\documentstyle[sprocl]{article}

\input{psfig.sty}
\input{epsf.sty}

\def\be{\begin{equation}}
\def\ee{\end{equation}}
\def\bea{\begin{eqnarray}}
\def\eea{\end{eqnarray}}
\def\lqcd{\Lambda_{\rm QCD}}
\def\mev{\,{\rm MeV}}
\def\gev{\,{\rm GeV}}
\def\case#1#2{\textstyle{{#1\over#2}}}
\def\lhqet{{\cal L}_{\rm HQET}}
\def\Dslash{\rlap{\,/}D}
\def\mbpole{m^{\rm pole}_b}
\def\mcpole{m^{\rm pole}_c}
\def\vckm{V_{\rm CKM}}
\def\hc{{\mbox{\rm h.c.}}}

\begin{document}

\begin{flushright}
JHU--TIPAC--200005\\
hep-ph/0007339\\
July, 2000
\end{flushright}

\vspace{1cm}

\title{THE CKM MATRIX AND THE HEAVY QUARK EXPANSION}

\author{ADAM F.~FALK}

\address{Department of Physics and Astronomy\\
The Johns Hopkins University\\
3400 North Charles Street, Baltimore, Maryland 21218} 

\maketitle\abstracts{These lectures contain an elementary introduction to heavy
quark symmetry and the heavy quark expansion.  Applications
such as the expansion of heavy meson decay constants and the
treatment of inclusive and exclusive semileptonic $B$ decays
are included.  The use of heavy quark methods for the
extraction of $|V_{cb}|$ and $|V_{ub}|$ is presented is some
detail.
}

\section{Heavy Quark Symmetry}

In these lectures I will introduce the ideas of heavy quark
symmetry and the heavy quark limit, which exploit the
simplification of certain aspects of QCD for infinite quark
mass, $m_Q\to\infty$.  We will see that while these ideas
are extraordinarily simple from a physical point of view,
they are of enormous practical utility in the study of the
phenomenology of bottom and charmed hadrons.  One reason for
this is the existence not just of an interesting new limit of
QCD, but of a systematic expansion about this limit.  The
technology of this expansion is the Heavy Quark Effective
Theory (HQET), which allows one to use heavy quark symmetry
to make accurate predictions of the properties and behavior
of heavy hadrons in which the theoretical errors are under
control.  While the emphasis in these lectures will be on the
physical picture of heavy hadrons which emerges in the heavy
quark limit, it will be important to introduce enough of the
formalism of the HQET to reveal the structure of the heavy
quark expansion as a simultaneous expansion in powers of
$\Lambda_{\rm QCD}/m_Q$ and
$\alpha_s(m_Q)$.  However, what I hope to leave you with
above all is an appreciation for the simplicity, elegance and
coherence of the ideas which underlie the technical results
which will be presented.  The interested reader is also
encouraged to consult a number of excellent
reviews,\cite{reviews} which typically cover in more detail
the material in these lectures.

\subsection{Introduction}

We begin by recalling the properties of charged current
interactions in the Standard Model.  They are mediated by
the interactions with the $W^\pm$ bosons, which for
the quarks take the form.
\be
  \pmatrix{\bar u&\bar c&\bar t}\gamma^\mu(1-\gamma^5)
  \vckm\pmatrix{d\cr s\cr b}W_\mu+\hc\,.
\ee
The $3\times3$ unitary matrix $\vckm$ is
\be
  \vckm=\pmatrix{V_{ud}&V_{us}&V_{ub}\cr
  V_{cd}&V_{cs}&V_{cb}\cr V_{td}&V_{ts}&V_{tb}\cr}\,.
\ee
The elements of $\vckm$ have a hierarchical structure, getting
smaller away from the diagonal: $V_{ud}$, $V_{cs}$ and
$V_{tb}$ are of order 1, $V_{us}$ and $V_{cd}$ are of order
$10^{-1}$, $V_{cb}$ and $V_{ts}$ are of order $10^{-2}$, and
$V_{ub}$ and $V_{td}$ are of order $10^{-3}$.  By contrast,
except for small effects associated with neutrino masses,
the interaction of the $W^\pm$ with the leptons is flavor
diagonal.

The CKM matrix is of fundamental importance, because it
is the low energy manifestation of the higher-energy physics
which breaks the global flavor symmetries of the Standard
Model.  In the absence of the Yukawa couplings, the
quark sector of the Standard Model may be characterized by its
gauge symmetry $SU(3)\times SU(2)\times U(1)$, and its
global symmetry $U(3)_Q\times U(3)_U\times U(3)_D$ which
rotates the triplets of colored fields $Q_L^i$, $U_R^i$ and
$D_R^i$ among each other.  The global symmetry group has a
total of $3\times 3^2=27$ generators.  With the addition of fields $\phi$
and $\tilde\phi$ (possibly composite) carrying Higgs and conjugate-Higgs
quantum numbers, one may write Yukawa-type interactions,
\be
  \lambda_u^{ij}\left(\overline{Q}_L^{\,i}\,\tilde\phi\,
  U_R^j\right)+
  \lambda_d^{ij}\left(\overline{Q}_L^{\,i}\,\phi\,
  D_R^j\right)+{\rm h.c.}
\ee
which break the flavor symmetries explicitly.  The
complex matrices $\lambda_u^{ij}$ and $\lambda_d^{ij}$
correspond to $2\times (2\times 3^2)=36$ independent
parameters.  They break the global flavor symmetries
completely, except for a remaining conserved baryon
number $U(1)_B$.  The $27-1=26$ broken generators may be
used to rotate 26 of the Yukawa couplings to zero, leaving
$36-26=10$ physical parameters.  

When $\phi$ and $\tilde\phi$ get vacuum expectation values,
one may go to the mass eigenstate basis for the quark fields,
in which case the ten parameters are six quark masses and
four parameters characterizing $\vckm$.  An independent
examination of $\vckm$ confirms this counting.  A $3\times3$
unitary matrix has 9 parameters, of which ${3\choose2}=3$ are
angles and the remaining 6 are complex phases.  However, one
may adjust 5 relative phases of the mass eigenstate quark
fields, leaving $6-5=1$ physical phase in $\vckm$.  Thus
$\vckm$ is indeed characterized by four parameters, one of
which is a CP-violating phase.

It is an instructive exercise, left to the reader, to
perform the analogous counting for the general case of
$U(N_f)_Q\times U(N_f)_U\times U(N_f)_D$ global flavor
symmetry.  One finds that the Yukawa couplings contain
$N_f^2+1$ physical parameters, of which $2N_f$ are quark
masses.  The remaining
$(N_f-1)^2$ parameters characterize $\vckm$, with
${1\over2}N_f(N_f-1)$ being angles and
${1\over2}(N_f-1)(N_f-2)$ being complex phases.  In
particular, one notes that for
$N_f=2$ there is no CP violation in the weak interactions.

Why is an understanding of QCD crucial to the study of the
properties of $\vckm$?  As an example,
consider semileptonic $b$ decay,
$b\to c\,\ell\bar\nu$, from which one would like to
extract $|V_{cb}|$.  This process is mediated by a
four-fermion operator,
\be\label{Obcdef}
  {\cal O}_{bc}={G_FV_{cb}\over\sqrt2}\bar c\gamma^\mu(1-\gamma^5)b\,
  \bar\nu\gamma_\mu(1-\gamma^5)\ell\,.
\ee
The weak matrix element is easy to calculate at the quark
level,
\be
  {\cal A}_{\rm quark} = 
  \langle c\,\ell\bar\nu|\,{\cal O}_{bc}|b\rangle=
  {G_FV_{cb}\over\sqrt2}\bar u_c(p_c)\gamma^\mu(1-\gamma^5)
  u_b(p_b)\,
  \bar u_\ell(p_\ell)\gamma_\mu(1-\gamma^5)v_\nu(p_\nu)\,.
\ee
However, ${\cal A}_{\rm quark}$ is only relevant at very
short distances; at longer distances, QCD confinement
implies that free $b$ and $c$ quarks are not asymptotic
states of the theory.  Instead, nonperturbative QCD effects
``dress'' the quark level transition $b\to c\,\ell\bar\nu$
to a hadronic transition, such as 
\be
  B\to D\ell\bar\nu\quad{\rm or}\quad B\to D^*\ell\bar\nu
  \quad{\rm or}\quad \ldots\,
\ee
(In these lectures, we will use a convention in which a $B$
meson contains a $b$ quark, not a $\bar b$ antiquark.)  The
hadronic matrix element 
${\cal A}_{\rm hadron}$ depends on nonperturbative QCD as
well as on $G_FV_{cb}$, and is difficult to calculate from
first principles.  To disentangle the weak interaction part
of this complicated process requires us to develop some
understanding of the strong interaction effects.

There are a variety of methods by which one can do this. 
Perhaps the most popular, historically, has been use of
various quark potential models.\cite{models}  While these
models are typically very predictive, they are based on
uncontrolled assumptions and approximations, and it is
virtually impossible to estimate the theoretical errors
associated with their use.  This is a serious defect if one
builds such a model into the experimental extraction of a
weak coupling constant such as $V_{cb}$, because the
uncontrolled theoretical errors then infect the experimental
result.

These are issues which are important for the extraction of
all the elements of $\vckm$.  Let us pause now to review our
current experimental knowledge of each of the magnitudes. 
The results are taken from the Particle Data Book.\cite{PDG} 
(The phases of the matrix elements must be extracted from CP
violating asymmetries, as discussed elsewhere at this school.)

We start with the submatrix describing mixing among
the first two generations.  The parameter $|V_{ud}|$ is
measured by studying the rates for neutron and nuclear
$\beta$ decay.  Here the isospin symmetry of the strong
interactions may be used to control the nonperturbative
dynamics, since the operator $\bar
d\gamma^\mu(1-\gamma^5)u$ which mediates the decay is a
partially conserved current associated with a generator of
chiral $SU(2)_L\times SU(2)_R$.  The current data yield
\be
  |V_{ud}|=0.9735\pm0.0008\,,
\ee
so $|V_{ud}|$ is known at the level of $0.1\%$.  The
parameter $|V_{us}|$ is measured similarly, via
$K\to\pi\ell\bar\nu_\ell$ and $\Lambda\to
p\ell\bar\nu_\ell$.  Here chiral $SU(3)L\times SU(3)_R$ must
be used in the hadronic matrix elements, since a strange
quark is involved; because the $m_s$ corrections are larger,
$|V_{us}|$ is only known to $1\%$:
\be
  |V_{us}|=0.2196\pm0.0023\,.
\ee

The $\vckm$ elements involving the charm quark are not so
well measured.  One way to extract $|V_{cs}|$ is to study
the decay $D\to K\ell^+\nu_\ell$.  Unfortunately, there is
no symmetry by which one can control the matrix element
$\langle K|\bar s\gamma^\mu(1-\gamma^5)c|D\rangle$, since
flavor $SU(4)$ is badly broken.  One is forced to resort to
models for these matrix elements.  The reported value is 
\be
  |V_{cs}|=1.04\pm0.16\,,
\ee
but it must be said that this error estimate is not on very
firm footing, and should probably be taken to be
substantially larger.

An alternative is to measure $V_{cs}$ from inclusive
processes at higher energies.  For example, one can study
the branching fraction for $W^+\to c\bar s$, which can be
computed using perturbative QCD.  The result of a
preliminary analysis is
\be
  |V_{cs}|=1.00\pm0.13\,,
\ee
consistent with the model-dependent measurement.  In this
case, however, the error is largely experimental, and is
unpolluted by hadronic physics.  Similarly, one extracts
$|V_{cd}|$ from deep inelastic neutrino scattering, using
the process $\nu_\mu+d\to c+\mu^-$.  This inclusive process
may be computed perturbatively in QCD, leading to a result
with accuracy at the level of~$10\%$,
\be
  |V_{cd}|=0.224\pm0.016\,.
\ee

The elements of $\vckm$ involving the third generation are,
for the most part, harder to measure accurately.  The
branching ratio for $t\to b\ell^+\nu$ can be analyzed
perturbatively, but the experimental data are not very
good.  The present bound on $|V_{tb}|$ is
\be
  {|V_{tb}|^2\over |V_{td}|^2+|V_{ts}|^2+|V_{tb}|^2}=
  0.99\pm0.29\,.
\ee
If one imposed the unitarity constraint
$|V_{td}|^2+|V_{ts}|^2+|V_{tb}|^2=1$, then this would amount
to a $15\%$ measurement of $|V_{tb}|$, but this unitarity
constraint is one of the properties of $\vckm$ which one is
trying to test.  More generally, in fact, one should be wary
of constraints on $\vckm$ which impose unitarity as part of
the analysis; while one often obtains tighter constraints in
this way, these constraints have a different meaning than do
direct measurements.  What the comparison of a constrained
and unconstrained ``measurement'' really tells you is
whether the direct determination may be used to test the
unitarity of $\vckm$.

Unfortunately, there are as yet no direct extractions of
$|V_{td}|$ or $|V_{ts}|$.  One often speaks of these
elements being measured in $B_d-\overline B_d$ and
$B_s-\overline B_s$ mixing, but again, the hypothesis that
the Standard Model is responsible for these processes is
something that one really wants to check.  The correct way
to view this part of the experimental program is to say that
the Standard Model, including the unitarity of $\vckm$,
constrains $V_{ts}$ and $V_{td}$ severely enough that
testable predictions can be made for the mixing parameters
$\Delta m_d$ and $\Delta m_s$.

This leaves us with the matrix elements $V_{ub}$ and
$V_{cb}$, for which we need an understanding of $B$ meson decay. 
In these lectures we will discuss an approach to understanding the
relevant hadronic physics which exploits the fact that the $b$ and
$c$ quarks are {\it heavy}, by which we mean that
$m_b,m_c\gg\lqcd$.  The scale $\lqcd$ is the typical energy at
which QCD becomes nonperturbative, and is of the order of hundreds
of MeV.  The physical quark masses are approximately
$m_b\approx4.8\gev$ and
$m_c\approx1.5\gev$.  The formalism which we will develop will
not make as many predictions as do potential models.  However,
the compensation will be that we will develop a {\it
systematic\/} expansion in powers of $\lqcd/m_{b,c}$, within
which we will be able to do concrete error analysis.  In
particular, we will be able to  estimate the error associated
with the fact that $m_c$ may not be very close to the asymptotic
limit $m_c\gg\lqcd$.  Even where this error may be substantial,
the fact that it is under control allows us to maintain
predictive power in the theory.

\subsection{The heavy quark limit}

Consider a hadron $H_Q$ composed of a heavy quark $Q$ and
``light degrees of freedom'', consisting of light quarks,
light antiquarks and gluons, in the limit $m_Q\to\infty$. 
The Compton wavelength of the heavy quark scales as the
inverse of the heavy quark mass, $\lambda_Q\sim1/m_Q$.  The
light degrees of freedom, by contrast, are characterized by
momenta of order $\lqcd$, corresponding to wavelengths
$\lambda_\ell\sim1/\lqcd$.  Since
$\lambda_\ell\gg\lambda_Q$, the light degrees of freedom
cannot resolve features of the heavy quark other than its
conserved gauge quantum numbers.  In particular, they cannot
probe the actual {\it value\/} of $\lambda_Q$, that is, the
value of $m_Q$.

We may draw the same conclusion in momentum space.  The structure
of the hadron $H_Q$ is determined by nonperturbative strong
interactions.  The asymptotic freedom of QCD implies that
when quarks and gluons exchange momenta $p$ much larger than
$\lqcd$, the process is perturbative in the strong coupling
constant $\alpha_s(p)$.  On the other hand, the typical
momenta exchanged by the light degrees of freedom with each
other and with the heavy quark are of order $\lqcd$, for
which a perturbative expansion is of no use.  For these
exchanges, however, $p<m_Q$, and the heavy quark $Q$ does not
recoil, remaining at rest in the rest frame of the hadron. 
In this limit, $Q$ acts as a static source of electric and
chromoelectric gauge field.  The chromoelectric field, which
holds $H_Q$ together, is nonperturbative in nature, but it is
independent of $m_Q$.  The result is that the properties of
the light degrees of freedom depend only on the presence of
the static gauge field, independent of the flavor and mass of
the heavy quark carrying the gauge charge.\footnote{Top
quarks decay too quickly for a static chromoelectric field to
be established around them, so the simplifications discussed
here are not relevant to them.}

There is an immediate implication for the spectroscopy of
heavy hadrons.  Since the interaction of the light degrees of
freedom with the heavy quark is independent of $m_Q$, then so
is the spectrum of their excitations.  It is these
excitations which determine the spectrum of heavy hadrons
$H_Q$.  Hence the {\it splittings\/} $\Delta_i\sim\lqcd$
between the various hadrons $H_Q^i$ are independent of $Q$
and, in the limit $m_Q\to\infty$, do not scale with $m_Q$.

\begin{figure}
\epsfysize7cm
\hfil\epsfbox{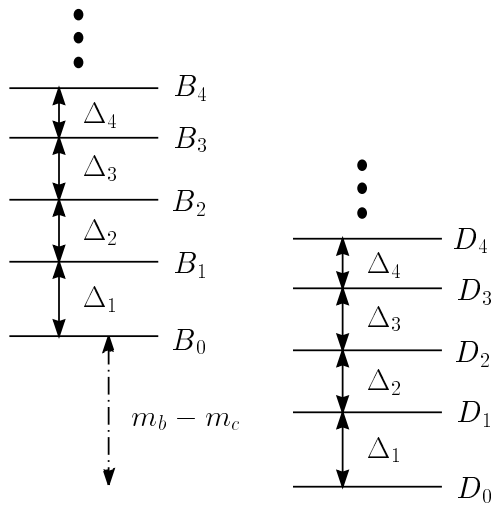}\hfill
\caption{Schematic spectra of the bottom and charmed mesons
in the limit $m_b,m_c\gg\lqcd$.  The offset of the two
spectra is not to scale; in reality,
$m_b-m_c\gg\Delta_i\sim\lqcd$.}
\label{fig:spectra}
\end{figure}

For example, the  bottom and charmed meson spectra are
shown schematically in Fig.~\ref{fig:spectra}, in the limit
$m_b,m_c\gg\lqcd$.  The light degrees of freedom are in
exactly the same state in the mesons $B_i$ and $D_i$, for a
given $i$.  The offset $B_i-D_i=m_b-m_c$ is just the
difference between the heavy quark masses; in no way does the
relationship between the spectra rely on an approximation
$m_b\approx m_c$.

We can enrich this picture by recalling that the heavy quarks
and light degrees of freedom also carry spin.  The heavy
quark has spin quantum number $S_Q={1\over2}$, which leads
to a chromomagnetic moment
\be
  \mu_Q={g\over2m_Q}\,.
\ee
Note that $\mu_Q\to0$ as $m_Q\to\infty$, and the interaction
between the spin of the heavy quark and the light degrees of
freedom is suppressed.  Hence the light degrees of freedom
are insensitive to $S_Q$; their state is independent of
whether $S_Q^z={1\over2}$ or $S_Q^z=-{1\over2}$.  Thus each
of the energy levels in Fig.~\ref{fig:spectra} is actually
doubled, one state for each possible value of  $S_Q^z$.

To summarize, what we see is that the light degrees of
freedom are the same when combined with any of the following
heavy quark states:
\be
  Q_1(\uparrow)\,,\quad Q_1(\downarrow);\quad 
  Q_2(\uparrow)\,,\quad Q_2(\downarrow);\quad\dots\quad
  Q_{N_h}(\uparrow)\,,\quad Q_{N_h}(\downarrow)\,,
\ee
where there are $N_h$ heavy quarks (in the real world,
$N_h=2$).  The result is an $SU(2N_h)$ symmetry which applies
to the light degrees of freedom.\cite{IW,VS,NW,PW,ES}  A new
symmetry means new nonperturbative relations between
physical quantities.  It is these relations which we wish to
understand and exploit.

The light degrees of freedom have total angular momentum
$J_\ell$, which is integral for baryons and half-integral for
mesons.  When combined with the  heavy quark spin
$S_Q={1\over2}$, we find physical hadron states with total
angular momentum
\be
  J=\left| J_\ell\pm\case12 \right|\,.
\ee
If $S_\ell\ne0$, then these two states are degenerate.  For
example, the lightest heavy mesons have $S_\ell=\case12$,
leading to a doublet with $J=0$ and $J=1$.  In the
charm system we find that the states of lowest mass are the
spin-0 $D$ and the spin-1 $D^*$; the corresponding bottom
mesons are the $B$ and $B^*$.  The heavy quark spin operator
$S_Q$ exchanges these two states.  Writing the spin wave
function $|m_Q, m_\ell\rangle$, we have
\be
  |M\rangle={1\over\sqrt2}\left(|\uparrow\downarrow\rangle
  -|\downarrow\uparrow\rangle\right),\qquad
  |M^*(J^3=0)\rangle=
  {1\over\sqrt2}\left(|\uparrow\downarrow\rangle
  +|\downarrow\uparrow\rangle\right).
\ee
Then it is easy to show that
\be
  S_Q|M\rangle={1\over2}|M^*(J^3=0)\rangle,\qquad
  S_Q|M^*(J^3=0)\rangle={1\over2}|M\rangle.
\ee

When effects of order $1/m_Q$ are included, the
chromomagnetic interactions split the states of given
$S_\ell$ but different $J$.  This ``hyperfine'' splitting is
not calculable perturbatively, but it is proportional to the
heavy quark magnetic moment $\mu_Q$.  This gives its scaling
with $m_Q$:
\bea
  m_{D^*}-m_D &\sim& 1/m_c\nonumber\\
  m_{B^*}-m_B &\sim& 1/m_b\,.
\eea
From this fact we can construct a relation which is a
nonperturbative prediction of heavy quark symmetry,
\be
  m_{B^*}^2-m_B^2=m_{D^*}^2-m_D^2\,.
\ee
Experimentally, $m_{B^*}^2-m_B^2=0.49\gev^2$ and
$m_{D^*}^2-m_D^2=0.55\gev^2$, so this prediction works quite
well.  Note that this relation involves not just the heavy
quark symmetry, but the systematic inclusion of the leading
symmetry violating effects.

Generally, the mass of a heavy hadron $H_Q$ may be expanded
in inverse powers of $m_Q$
\be
  m(H_Q)=m_Q+\bar\Lambda_H+{\cal O}(1/m_Q)\,,
\ee
where $\bar\Lambda_H$ is independent of $Q$ and is
associated with the energy of the light degrees of freedom in the
hadron $H_Q$. For the lowest lying $J_\ell={1\over2}$ doublet,
this quantity is usually just referred to as $\bar\Lambda$. 
From dimensional considerations, one expects
$\bar\Lambda$ to be of order of a few hundred MeV.

So far, we have formulated heavy quark symmetry for hadrons
in their rest frame.  Of course, we can easily boost to a
frame in which the hadrons have arbitrary four-velocity
$v^\mu=\gamma(1,\vec v)$.  For heavy quarks $Q_1$ and $Q_2$,
the symmetry will then relate hadrons $H_1(v)$ and $H_2(v)$
with the same velocity but with different momenta.  This
distinguishes heavy quark symmetry from ordinary symmetries
of QCD, which relate states of the same momentum.  To remind
ourselves of this distinction, henceforth we will label
heavy hadrons explicitly by their velocity: $D(v)$,
$D^*(v)$, $B(v)$, $B^*(v)$, and so on.

\subsection{Semileptonic decay of a heavy quark}

Now let us return to the semileptonic weak decay $b\to
c\,\ell\bar\nu$, but now consider it in the heavy quark limit
for the $b$ and $c$ quarks.  Suppose the decay occurs at time
$t=0$.  For $t<0$, the $b$ quark is embedded in a hadron
$H_b$; for $t>0$, the $c$ quark is dressed by light degrees
of freedom to $H_c'$.  Let us consider the lightest hadrons,
$H_b=B(v)$ and $H_c'=D(v')$.  Note that since the leptons
carry away energy and momentum, in general $v\ne v'$.

What happens to the light degrees of freedom when the heavy
quark decays?  For $t<0$, they see the chromoelectric field
of a point source with velocity $v$.  At $t=0$, this point
source recoils instantaneously\footnote{The weak decay
occurs over a very short time $\delta
t\sim1/M_W\ll1/\lqcd$.} to velocity $v'$; the color neutral
leptons do not interact with the light hadronic degrees of
freedom as they fly off.  The light quarks and gluons then
must reassemble themselves about the recoiling color
source.  This nonperturbative process will generally involve
the production of an excited state or of additional
particles; the light degrees of freedom can exchange energy
with the heavy quark, so there is no kinematic restriction on
the excitations (of energy $\sim\lqcd$) which can be formed. 
There is also some chance that the light degrees of freedom
will reassemble themselves back into a ground state $D$
meson.  The amplitude for this to happen is a function only
of the inner product $w=v\cdot v'$ of the initial and final
velocities of the color sources.  This amplitude, $\xi(w)$,
is known as the Isgur-Wise function.\cite{IW}

Clearly, the kinematic point $v=v'$, or $w=1$, is a special
one.  In this corner of phase space, where the leptons are
emitted back to back, there is no recoil of the source of
color field at $t=0$.  As far as the light degrees of freedom
are concerned, {\it nothing happens!\/}  Their state is
unaffected by the decay of the heavy quark; they don't even
notice it.  Hence the amplitude for them to remain in the
ground state is exactly unity.  This is reflected in a
nonperturbative normalization of the Isgur-Wise function at
zero recoil,\cite{IW}
\be\label{norm}
  \xi(1)=1\,.
\ee
As we will see, this normalization condition is of enormous
phenomenological use.  It will be extremely important to
understand the corrections to this result for finite heavy
quark masses $m_b$ and, especially, $m_c$.

The weak decay $b\to c$ is mediated by a left-handed current 
$\bar c\gamma^\mu(1-\gamma^5)b$.  Not only does this
operator carry momentum, but it can change the orientation of
the spin $S_Q$ of the heavy quark during the decay.  For a
fixed light angular momentum $J_\ell$, the relative
orientation of $S_Q$ determines whether the physical hadron
in the final state is a $D$ or a $D^*$.  However, the light
degrees of freedom are insensitive to $S_Q$, so the
nonperturbative part of the transition is the same whether it
is a $D$ or a $D^*$ which is produced.  Hence heavy quark
symmetry implies relations between the hadronic matrix
elements which describe the semileptonic decays $B\to
D\ell\bar\nu$ and $B\to D^*\ell\bar\nu$.

It is conventional to parameterize these matrix elements by a
set of scalar form factors.  These are defined separately for
the vector and axial currents, as follows:
\bea\label{formfactors}
  \langle D(v')|\,\bar c\gamma^\mu b\,|B(v)\rangle &=&
  h_+(w)(v+v')^\mu+h_-(w)(v-v')^\mu\nonumber\\
  \langle D^*(v',\epsilon)|\,\bar c\gamma^\mu b\,|B(v)\rangle 
  &=&
  h_V(w)i\varepsilon^{\mu\nu\alpha\beta}
  \epsilon_\nu^*v'_\alpha v_\beta
  \nonumber\\
  \langle D(v')|\,\bar c\gamma^\mu\gamma^5 b\,|B(v)\rangle 
  &=& 0\\
  \langle D^*(v',\epsilon)|\,\bar c\gamma^\mu\gamma^5 b\,
  |B(v)\rangle &=&
  h_{A_1}(w)(w+1)\epsilon^{*\mu}-\epsilon^*\cdot v
  [h_{A_2}(w)v^\mu+h_{A_3}(w)v^{\prime\mu}]\,.\nonumber
\eea
The set of form factors $h_i(w)$ is the one appropriate to
the heavy quark limit.  Other linear combinations are also
found in the literature.  In any case, the form factors are
independent nonperturbative functions of the recoil or
equivalently, for fixed $m_b$ and $m_c$, of the momentum
transfer.  However, in the heavy quark limit they correspond
to a {\it single\/} transition of the light degrees of
freedom, being distinguished from each other only by the
relative orientation of the spin of the heavy quark.  Hence
they may all be written in terms of the single function
$\xi(w)$ which describes this nonperturbative transition. 
As we will derive later, the result is a set of
relations,\cite{IW}
\bea\label{ffrelations}
  &&h_+(w)=h_V(w)=h_{A_1}(w)=h_{A_3}(w)=\xi(w)\nonumber\\
  &&h_-(w)=h_{A_2}(w)=0\,,
\eea
which follow solely from the heavy quark symmetry.  Of
course, all of the form factors which do not vanish inherit
the normalization condition (\ref{norm}) at zero recoil. 
This result is a powerful constraint on the structure of
semileptonic decay in the heavy quark limit.

\subsection{Heavy meson decay constant}
\label{sec:HQSf}

As a final example of the utility of the heavy quark limit,
consider the coupling of the heavy meson field to the axial
vector current.  This is conventionally parameterized by of a
decay constant; for example, for the $B^-$ meson we define $f_B$
via
\be\label{fBdef}
  \langle 0|\,\bar u\gamma^\mu\gamma^5 b\,|B^-(p_B)\rangle
  =if_Bp_B^\mu\,.
\ee
What is the dependence of the nonperturbative quantity $f_B$
on $m_B$?  To address this question, we rewrite
Eq.~(\ref{fBdef}) in a form appropriate to taking the heavy
quark limit, $m_B\to\infty$ (which is equivalent to
$m_b\to\infty$).  This entails making explicit the
dependence of all quantities on $m_B$.  First, we trade the
$B^-$ momentum for its velocity,
\be
  p_B^\mu=m_B v^\mu\,.
\ee
Second, we replace the usual $B^-$ state, whose normalization
depends on $m_B$,
\be
  \langle B(p_1)|B(p_2)\rangle=2E_B\,\delta^{(3)}
  (\vec{p}_1-\vec{p}_2)\,,
\ee
by a mass-independent state,
\be
  |B(v)\rangle={1\over\sqrt{m_B}}\,|B(p_B)\rangle\,,
\ee
satisfying
\be
  \langle B(v_1)|B(v_2)\rangle=2\gamma\,\delta^{(3)}
  (\vec{p}_1-\vec{p}_2)\,.
\ee
Then Eq.~(\ref{fBdef}) becomes
\be
  \sqrt{m_B}\langle 0|\,\bar u\gamma^\mu\gamma^5 b\,
  |B^-(v)\rangle
  =if_Bm_Bv^\mu\,.
\ee
The nonperturbative matrix element $\langle 0|\,
\bar u\gamma^\mu\gamma^5 b\,|B^-(v)\rangle$ is independent
of $m_B$ in the heavy quark limit.  Hence, we see that in
this limit $f_B$ takes the form
\be
  f_B=m_B^{-1/2}\times{\rm (independent\ of\ }m_B{\rm )}\,.
\ee
This makes explicit the scaling of $f_B$ with $m_B$.  It is
more interesting to write this as a prediction for the ratio
of charmed and bottom meson decay constants.  We
find~\cite{IW,VS,PW}
\be\label{fBfD}
  {f_B\over f_D}=\sqrt{m_D\over m_B} 
  + O\left({\lqcd\over m_D},{\lqcd\over m_B}\right)\,.
\ee
For the physical bottom and charm masses, of course, the
correction terms proportional to $\lqcd/m_Q$ could be
important.

\section{Effective Field Theories}
\label{Chap2:secAA}

\subsection{General Considerations}
\label{Chap2:ssecAA}

A central observation which underlies much of the theoretical study of
$B$ mesons is that physics at a wide variety of distance (or momentum) scales is typically
relevant in a given process.  At the same time, the physics at different scales must often
be analyzed with different theoretical approaches.  Hence it is crucial to have a tool
which enables one to identify the physics at a given scale and to separate it out
explicitly.  Such a tool is the operator product expansion, used in  conjunction with the
renormalization group.  Here a general discussion of its
application is given.

\begin{figure}
\epsfysize4cm
\hfil\epsfbox{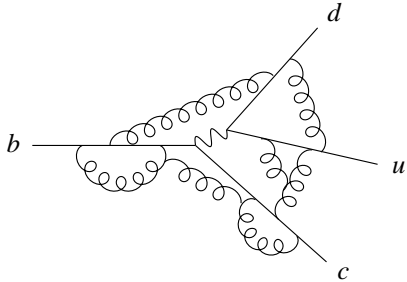}\hfill
\caption{The nonleptonic decay of a $b$ quark.}
\label{Chap2:Chap2fig1}
\end{figure}

Consider the Feynman diagram shown in Fig.~\ref{Chap2:Chap2fig1}, in which a 
$b$ quark decays nonleptonically.  The virtual quarks and gauge bosons have virtualities
$\mu$ which vary widely, from $\Lambda_{\rm QCD}$ to $M_W$ and higher.  Roughly speaking,
these virtualities can be classified into a variety of energy regimes: (i) $\mu\gg M_W$;
(ii) $M_W\gg\mu\gg m_b$; (iii) $m_b\gg\mu\gg\Lambda_{\rm QCD}$; (iv)
$\mu\approx\Lambda_{\rm QCD}$.  Each of these momenta corresponds to a different distance
scale; by the uncertainty principle, a particle of virtuality $\mu$ can propagate a
distance $x\approx 1/\mu$ before being reabsorbed.  At a given resolution $\Delta x$, only
some of these virtual particles can be distinguished, namely those that propagate a
distance $x>\Delta x$.  For example, if $\Delta x>1/M_W$, then the virtual $W$ cannot be
seen, and the process whereby it is exchanged would appear as a point interaction.  By the
same token, as
$\Delta x$ increases toward $1/\Lambda_{\rm QCD}$, fewer and fewer of the virtual gluons
can be seen explicitly.  Finally, for
$\mu\approx\Lambda_{\rm QCD}$, it is inappropriate to speak of
virtual gluons at all, because at such low momentum scales QCD
becomes strongly interacting and an expansion in
terms of individual gluons is inadequate.

It is useful to organize the computation of a diagram such as is shown in
Fig.~\ref{Chap2:Chap2fig1} in terms of the  virtuality of the exchanged particles.  This
is important both conceptually and practically.  First, it is often the case that a
distinct set of approximations and approaches is useful at each distance scale, and one 
would like to be able to apply specific theoretical techniques at the scale at which they
are appropriate.  Second, Feynman diagrams in which two distinct scales $\mu_1\gg\mu_2$
appear together can lead to logarithmic corrections of the form
$\alpha_s^n\ln^n(\mu_1/\mu_2)$, which for
$\ln(\mu_1/\mu_2)\sim1/\alpha_s$ can spoil the perturbative
expansion.  A proper separation of scales will include a
resummation of such terms.

\subsection{Example I:  Weak $b$ Decays}
\label{Chap2:ssecAB}

As an example, consider the weak decay of a $b$ quark, $b\to c\bar u d$, which is mediated
by the decay of a virtual $W$ boson.  Viewed with resolution $\Delta x<1/M_W$, the decay
amplitude involves an explicit
$W$ propagator and is proportional to
\begin{equation}
\label{Chap2:eq:matel1}
  \bar c\gamma^\mu(1-\gamma^5)b\,\bar d\gamma_\mu(1-\gamma^5)u
  \times{(ig_2)^2\over p^2-M_W^2}\,,
\end{equation} where $p^\mu$ is the momentum of the virtual $W$.  Since $m_b\ll M_W$, the
kinematics constrains $p^2\ll M_W^2$, so the virtuality of the $W$ is of order $M_W$, and
it travels a distance of order $1/M_W$ before decaying.  Viewed with a lower resolution,
$\Delta x>1/M_W$, the process $b\to c\bar u d$ appears to be a local interaction, with four
fermions interacting via a potential which is a $\delta$ function where the four particles
coincide.  This can be seen by making a Taylor expansion of the amplitude in powers of
$p^2/M_W^2$,
\begin{equation}
  \bar c\gamma^\mu(1-\gamma^5)b\,\bar d\gamma_\mu(1-\gamma^5)u
  \times{g_2^2\over M_W^2}\left[1+{p^2\over M_W^2}
  +{p^4\over M_W^4}+\dots\right]\,.
\end{equation} The coefficient of the first term is just the usual Fermi decay constant,
$G_F/\sqrt2$.  The higher order terms correspond to local operators of higher mass
dimension.  In the sense of a Taylor expansion, the momentum-dependent matrix element
(\ref{Chap2:eq:matel1}), which involves the propagation of a $W$ boson between {\it two\/}
spacetime points, is identical to the matrix element of the following infinite sum of
local operators:
\begin{equation}
\label{Chap2:eq:nonlepeff}
  {G_F\over\sqrt2}\,\bar c\gamma^\mu(1-\gamma^5)b
  \left[1+{(i\partial)^2\over M_W^2}+{(i\partial)^4\over M_W^4}
  +\dots\right]
  \bar d\gamma_\mu(1-\gamma^5)u\,,
\end{equation} where the derivatives act on the entire current on the right.  This
expansion of the nonlocal product of currents in terms of local operators,  sometimes
known as an {\it operator product expansion,} is valid so long as  
$p^2\ll M_W^2$.  For $B$ decays, the external kinematics requires 
$p^2\le m_b^2$,  so this condition is well satisfied.  In this regime, one may consider a
nonrenormalizable {\it effective field theory,} with interactions of dimension six and
above.  The construction of such a low energy effective theory is  also known as {\it
matching.}  As it is nonrenormalizable, the effective theory is defined (by construction)
only up to a cutoff, in this case $M_W$.  The cutoff is explicitly the mass of a particle
which has been removed from the theory, or {\it integrated out.}  If one considers
processes  in which one is restricted kinematically to momenta well below the cutoff, the
nonrenormalizability of the theory poses no technical problems.  Although the coefficients
of operators of dimension greater than six require counterterms  in the effective theory
(which may be unknown in strongly interacting  theories),   their matrix elements are
suppressed by  powers of
$p^2/M_W^2$.  To a {\it given order\/} in $p^2/M_W^2$, the theory is well-defined and
predictive.

From a modern point of view, in fact, such nonrenormalizable effective theories are
actually preferable to renormalizable theories, because the nonrenormalizable terms
contain information about the energy  scale at  which the theory ceases to apply.  By
contrast, renormalizable  theories contain no such explicit clues about their region of
validity.

In principle, it is possible to include effects beyond leading order in
$p^2/M_W^2$ in the effective theory, but in practice, this is usually quite complicated
and rarely worth the effort.  Almost always, the operator product expansion is truncated
at dimension six, leaving only the four-fermion contact term.  Corrections to this
approximation are of order
$m_b^2/M_W^2\sim10^{-3}$.

\subsection{Radiative Corrections}

\begin{figure}
\epsfysize4cm
\hfil\epsfbox{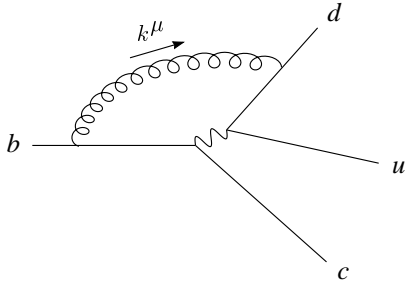}\hfill
\caption{The nonleptonic decay of a $b$ quark at one loop.}
\label{Chap2:Chap2fig2}
\end{figure}

At tree level, the effective theory is constructed simply by integrating out the $W$
boson, because this is the only particle in a tree level diagram which is off-shell by
order $M_W^2$.  When radiative corrections are included, gluons and light quarks can also
be off-shell by this order.  Consider the one-loop diagram shown in
Fig.~\ref{Chap2:Chap2fig2}.   The components of the loop momentum $k^\mu$ are allowed to
take all values in the loop integral.  However, the integrand is cut off both in the
ultraviolet and in the infrared.  For $k>M_W$, it scales as $d^4k/k^6$, which is
convergent as $k\to\infty$.  For $k<m_b$, it scales as $d^4k/k^3m_bM_W^2$, which is
convergent as $k\to0$.  In between, all momenta in the range
$m_b<k<M_W$ contribute to the integral with roughly equivalent weight. 

As a consequence, there is potentially a radiative correction
proportional to $\alpha_s\ln(M_W/m_b)$.  Even if
$\alpha_s(\mu)$ is evaluated at the high scale $\mu=M_W$, such
a term is not small in the limit $M_W\to\infty$.  At
$n$ loops, there is potentially a term of order
$\alpha_s^n\ln^n(M_W/m_b)$.  For $\alpha_s\ln(M_W/m_b)\sim1$,
these terms need to be resummed for the perturbation series to
be predictive.  The technique for performing such a
resummation is the {\it renormalization group.}

The renormalization group exploits the fact that in the effective theory, operators such
as 
\begin{equation}
\label{Chap2:eqI}
  O_I=\bar c_i\gamma^\mu(1-\gamma^5)b^i\,
  \bar d_j\gamma_\mu(1-\gamma^5)u^j
\end{equation}   receive radiative corrections and must be subtracted and renormalized. 
(Here the color indices $i$ and 
$j$ are explicit.)  In dimensional regularization, this means that they acquire, in 
general,  a dependence on the renormalization scale $\mu$.  Because physical  predictions
are independent of $\mu$, in the renormalized effective theory
the operators must be multiplied by coefficients with a
compensating dependence on $\mu$.  It is also possible for
operators to mix under renormalization, so the set of operators
induced at tree level may be enlarged once radiative corrections
are included.  In the present example, a second operator with
different color structure,
\begin{equation}
\label{Chap2:eqII}
  O_{II}=\bar c_i\gamma^\mu(1-\gamma^5)b^j\,
  \bar d_j\gamma_\mu(1-\gamma^5)u^i\,,
\end{equation} is induced at one loop. The interaction Hamiltonian of the effective theory
is then
\begin{equation}
\label{Chap2:eq:Heff}
  {\cal H}_{\rm eff}=C_I(\mu)O_I(\mu)+C_{II}(\mu)O_{II}(\mu)\,,
\end{equation} and it satisfies the differential equation
\begin{equation}
  \mu{d\over d\mu}{\cal H}_{\rm eff}=0\,.
\end{equation} By computing the dependence on $\mu$ of the operators $O_i(\mu)$, one can
deduce the $\mu$-dependence of the {\it Wilson coefficients\/} 
$C_i(\mu)$.   In this case, a simple calculation yields
\begin{equation}
\label{Chap2:eq:lowil}
  C_{I,II}(\mu)=
  {1\over2}\left[\left({\alpha_s(M_W)\over\alpha_s(\mu)}\right)^{6/23}
  \pm\left({\alpha_s(M_W)\over\alpha_s(\mu)}\right)^{-12/23}\right]\,.
\end{equation} For $\mu=m_b$, these expressions resum all large logarithms proportional to
$\alpha_s^n\ln^n(M_W/m_b)$.

The decays which are observed involve physical hadrons, not asymptotic quark states.  For
example, this nonleptonic $b$ decay can be realized in the channels $B\to D\pi$, $B\to
D^*\pi\pi$, and so on.  The computation of partial decay rates for such processes requires
the analysis of hadronic matrix  elements such as 
\begin{equation}
  \langle D\pi|\,\bar c\gamma^\mu(1-\gamma^5)b\,
  \bar u\gamma_\mu(1-\gamma^5)d\,|\bar b\rangle\,.
\end{equation} Such matrix elements involve nonperturbative QCD and are extremely
difficult to compute from first principles.  However, they have no intrinsic dependence on
large mass scales such as $M_W$.  Because of this, they should naturally be evaluated at a
renormalization scale $\mu\ll M_W$, in which case large  logarithms 
$\ln(M_W/m_b)$ will not arise in the matrix elements.    By choosing such a  low scale in
the effective theory (\ref{Chap2:eq:Heff}),  all such terms are resummed into the
coefficient functions $C_i(m_b)$.  As promised, the physics at scales near $M_W$ has been
separated from the physics at scales near
$m_b$, with the renormalization group used to resum the large logarithms which connect
them. In fact, as we will see in the next section, nonperturbative
hadronic matrix elements are usually evaluated at an even lower
scale $\mu\approx\Lambda_{\rm QCD}\ll m_b$, explicitly resumming
all perturbative QCD corrections.

\section{Heavy Quark Effective Theory}

We have already extracted quite a bit of nontrivial
information from the heavy quark limit.  We have found the
scaling of various quantities with $m_Q$, we have studied the
implications for heavy hadron spectroscopy, and we have found
nonperturbative relations among the hadronic form factors
which describe semileptonic $b\to c$ decay.  However, all of
these results have been obtained in the strict limit
$m_Q\to\infty$.  If the heavy quark limit is to be of more
than academic interest, and is to provide the basis for
quantitative phenomenology, we have to understand how to
include corrections systematically.  There are actually two
types of corrections which we would like to include.  {\it
Power corrections\/} are subleading terms in the expansion
in $\lqcd/m_Q$; those proportional to $\lqcd/m_c$ are the
most worrisome, because of the relatively small charm quark
mass.  {\it Logarithmic corrections\/} arise from the
implicit dependence of quantities on $m_Q$ through the strong
coupling constant $\alpha_s(m_Q)\sim1/\ln(m_Q/\lqcd)$.  For
the physical values of $m_b$ and $m_c$, either of these could
be important.  What we need is a formalism which can
accommodate them both.

In short, we need to go from a set of heavy quark symmetry
predictions in the
$m_Q\to\infty$ limit, to a reformulation of QCD which
provides a controlled expansion about this limit.  The
formalism which does the job is the Heavy Quark Effective
Theory, or the HQET.  The purpose of the HQET is to allow us
to extract, explicitly and systematically, all dependence of
physical quantities on $m_Q$, in the limit $m_Q\gg\lqcd$.  In
these lectures, we will develop only enough of the
technology to treat the dominant leading effects, providing
indications along the way of how one would carry the
expansion further.

The HQET, as formulated here, was developed in a series of
papers going back to the late
1980's,\cite{IW,VS,PW,EH,Georgi,Grin,FGGW,FG,Luke,FGL,MRR}
which the reader who is interested in tracking its historical
development may consult.

\subsection{The effective Lagrangian}

Consider the kinematics of a heavy quark $Q$, bound in a
hadron with light degrees of freedom to make a color singlet
state.  The small momenta which $Q$ typically exchanges with
the rest of the hadron are of order $\lqcd\ll m_Q$, and they
never take $Q$ far from its mass shell,
$p_Q^2=m_Q^2$.  Hence the momentum $p_Q^\mu$ can be
decomposed into two parts,
\be
  p_Q^\mu=m_Q v^\mu+k^\mu\,,
\ee
where $m_Qv^\mu$ is the constant on-shell part of $p_Q^\mu$,
and $k^\mu\sim\lqcd$ is the small, fluctuating ``residual
momentum''.  The on-shell condition for the heavy quark then
becomes
\be
  m_Q^2=(m_Qv^\mu+k^\mu)^2=m_Q^2+2m_Qv\cdot k+k^2\,.
\ee
In the heavy quark limit we may neglect the last term compared to
the second, and we have the simple condition
\be
  v\cdot k=0
\ee
for an on-shell heavy quark.  Here the velocity $v^\mu$
functions as a label; since soft interactions cannot change
$v^\mu$, there is a {\it velocity superselection rule\/} in
the heavy quark limit, and $v^\mu$ is a good quantum number
of the QCD Hamiltonian.

We find the same result by taking the $m_Q\to\infty$ limit of
the heavy quark propagator,
\be
  {i\over\rlap/p-m_Q+i\epsilon}\to{1+\rlap/v\over2}\,
  {i\over v\cdot k+i\epsilon}\,.
\ee
In this limit the propagator is independent of $m_Q$.  The
projection operators
\be
  P_\pm={1\pm\rlap/v\over2}
\ee
project onto the positive ($P_+$) and negative ($P_-$)
frequency parts of the Dirac field $Q$.  This is clear in the
Dirac representation in the rest frame, in which $P_+$ and
$P_-$ project, respectively, onto the upper two and lower two
components of the heavy quark spinor.  In the limit
$m_Q\to\infty$, in which $Q$ remains almost on shell, only
the ``large'' upper components of the field $Q$ propagate;
mixing via zitterbewegung with the ``small'' lower
components is suppressed by $1/2m_Q$.  Hence the action of
the projectors on $Q$ is
\be
  P_+Q(x)=Q(x)+O(1/m_Q)\,,\qquad P_-Q(x)=0+O(1/m_Q)\,.
\ee
More precisely, these relations should be understood as
pertaining to those modes of the field $Q(x)$ which
annihilate heavy quarks and antiquarks in a heavy meson.

The momentum dependence of the field $Q$ is given by its
action on a heavy quark state,
\be 
  Q(x)\,|Q(p)\rangle=e^{-ip\cdot x}\,|0\rangle\,.
\ee
If we now multiply both sides by a phase corresponding to the
on-shell momentum,
\be 
  e^{im_Qv\cdot x}Q(x)\,|Q(p)\rangle=e^{-ik\cdot x}\,
  |0\rangle\,,
\ee
the right side of this equation is independent of $m_Q$. 
Hence the left side must be, as well.  Combining this
observation with the argument of the previous paragraph, we
are motivated to define a $m_Q$-{\it independent\/}
effective heavy quark field $h_v(x)$,
\be
  h_v(x)=e^{im_Qv\cdot x}\,P_+\,Q(x)\,.
\ee
Note that the effective field carries a velocity label $v$
and is a two-component object.  The modifications to the
ordinary field $Q(x)$ project out the positive frequency part
and ensure that states annihilated by $h_v(x)$ have no
dependence on $m_Q$.  Hence, these are reasonable candidate
fields to carry representations of the heavy quark symmetry.
Of course, the small components cannot be neglected when
effects of order $1/m_Q$ are included.  In the HQET they are
represented by a field
\be
  H_v(x)=e^{im_Qv\cdot x}\,P_-\,Q(x)\,.
\ee
The field $H_v(x)$ vanishes in the $m_Q\to\infty$ limit.

The ordinary QCD Lagrange density for a field $Q(x)$ is given
by
\be
  {\cal L}_{\rm QCD}=\overline Q(x)\,(i\Dslash-m_Q)\,Q(x)\,,
\ee
where $D_\mu=\partial_\mu-igA^a_\mu T^a$ is the gauge
covariant derivative.  To find the Lagrangian of the HQET, we
substitute
\be
  Q(x)=e^{-im_Qv\cdot x}h_v(x)+\dots
\ee
into ${\cal L}_{\rm QCD}$ and expand.  With the aid of the
projection identity $P_+\gamma^\mu P_+=v^\mu$, we find
\be
  \lhqet= \bar h_v(x)iv\cdot Dh(x)\,.
\ee
This simple Lagrangian leads to the propagator we derived
earlier,
\be
  {i\over v\cdot k+i\epsilon}\,,
\ee
and to an equally simple quark-gluon vertex,
\be
  igT^av^\mu A^a_\mu\,.
\ee
Note that both the propagator and the vertex are independent
of $m_Q$, reflecting the heavy quark flavor symmetry.  They
also have no Dirac structure, reflecting the heavy quark
spin symmetry.  Our intuitive statements about the structure
of heavy hadrons have been promoted to explicit symmetries of
the QCD Lagrangian in the limit $m_Q\to\infty$.

It is straightforward to include power corrections to
$\lhqet$.  Write $Q(x)$ in terms of the effective fields,
\be
  Q(x)=e^{-im_Qv\cdot x}\left[h_v(x)+H_v(x)\right]\,,
\ee
and apply the classical equation of motion
$(i\Dslash-m_Q)Q(x)=0$:
\be
  i\Dslash\,h_v(x)+(i\Dslash-2m_Q)H_v(x)=0\,.
\ee
Multiplying by $P_-$ and commuting $\rlap/v$ to the right, we
find
\be
  (iv\cdot D+2m_Q)H_v(x)=i\Dslash_\perp\,h_v(x)\,,
\ee
where $D^\mu_\perp=D^\mu-v^\mu v\cdot D$.  We then
substitute $Q(x)$ into ${\cal L}_{\rm QCD}$ as before,
eliminate $H_v(x)$ and expand in $1/m_Q$ to obtain
\bea\label{hqetlagrangian}
  \lhqet&=&\bar h_viv\cdot Dh_v+\bar h_vi\Dslash_\perp
  \,{1\over iv\cdot D+2m_Q}\,i\Dslash_\perp h_v\nonumber\\
  &=&\bar h_viv\cdot Dh_v+{1\over2m_Q}
  \left[\bar h_v(iD_\perp)^2h_v
  +{g\over2}\,\bar h_v\sigma^{\alpha\beta}
  G_{\alpha\beta}h_v\right]
  +\dots\,.
\eea
The leading corrections have a simple interpretation, which
becomes clear in the rest frame, $v^\mu=(1,0,0,0)$.  The
spin-independent term is
\be
  {1\over2m_Q}\,{\cal O}_K={1\over2m_Q}\,
  \bar h_v(iD_\perp)^2h_v\to 
  -{1\over2m_Q}\,\bar h_v(i\vec D)^2h_v\,,
\ee
which is the negative of the nonrelativistic kinetic energy
of the heavy quark.  Because of the explicit factor of
$1/2m_Q$, this term violates the heavy flavor symmetry.  The
spin-dependent part is
\be
  {1\over2m_Q}\,{\cal O}_G=
  {1\over2m_Q}\,{g\over2}\,\bar h_v\sigma^{\alpha\beta}
  G_{\alpha\beta}h_v\to
  {1\over4m_Q}\,\bar h_v\sigma^{ij}T^ah_v\times gG_{ij}^a=
  g\vec\mu_Q^{\,a}\cdot\vec B^{\,a},
\ee
which is the coupling of the spin of the heavy quark to the
chromomagnetic field.  Because it has a nontrivial Dirac
structure, this term violates both the heavy flavor symmetry
and the heavy spin symmetry.  For example, ${\cal O}_G$ is
responsible for the $D-D^*$ and $B-B^*$ mass splittings. 
These correction terms will be treated as part of the {\it
interaction\/} Lagrangian, even though ${\cal O}_K$ has a
piece which is a pure bilinear in the heavy quark field.

\subsection{Effective currents and states}

The expansion of the weak interaction current $\bar
c\gamma^\mu(1-\gamma^5)b$ is analogous.  However, here we
must introduce separate effective fields for the charm and
bottom quarks, each with its own velocity:
\be
  b\to h^b_v\,,\qquad c\to h^c_{v'}\,.
\ee
Then a general flavor-changing current becomes, to leading
order,
\be
  \bar c\,\Gamma\, b\to\bar h^c_{v'}\,\Gamma\, h^b_v\,,
\ee
where $\Gamma$ is a fixed Dirac structure.  With the leading
power corrections, this is
\be\label{current}
  \bar c\,\Gamma\, b\to\bar h^c_{v'}\,\Gamma\, h^b_v
  +{1\over2m_b}\,\bar h^c_{v'}\Gamma(i\Dslash_\perp)h^b_v
  +{1\over2m_c}\,\bar h^c_{v'}(-i\overleftarrow\Dslash_\perp)
  \Gamma h^b_v
  +\dots\,.
\ee
The effective currents, and other operators which appear in
the HQET, may often be simplified by use of the classical
equation of motion,
\be
  iv\cdot D h_v(x)=0\,.
\ee
However, it is only safe to apply these equations naively at
order $1/m_Q$; at higher order the application of the
equations of motion involves additional
subtleties.\cite{Pol,FN,FLS94}

To complete the effective theory, we need $m_Q$-independent
hadron states which are created and annihilated by currents
containing the effective fields.  For example, there is an
effective pseudoscalar meson state $|M(v)\rangle$ which
couples to the effective axial current $\bar
q\gamma^\mu\gamma^5 h_v$, with a coupling $F_M$ which is
independent of $m_Q$,
\be\label{matfm}
  \langle0|\,\bar q\gamma^\mu\gamma^5 h_v\,|M(v)\rangle
  =iF_Mv^\mu\,.
\ee
At lowest order, $F_M$ is related to a conventional decay
constant such as $f_B$ by 
\be
  f_B=F_M/\sqrt{m_B}\,,
\ee
from which we immediately find the relationship (\ref{fBfD})
between $f_D$ and $f_B$.

\begin{figure}
\epsfysize2.5cm
\hfil\epsfbox{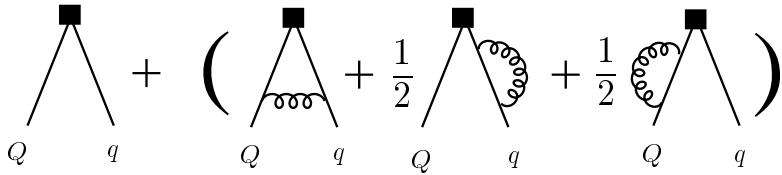}\hfill
\caption{Tree level plus the one loop renormalization of the
current $\bar q\Gamma Q$ in QCD.  The box represents the
current insertion.}
\label{fig:fqcd}
\end{figure}

\subsection{Radiative corrections}

We can use the effective Lagrangian $\lhqet$ to compute the
radiative corrections to the matrix element (\ref{matfm}). 
In particular, we would like to extract the dependence of
$F_M$ on $\ln m_Q$.  This dependence comes through the
one-loop renormalization of the current $\bar
q\gamma^\mu\gamma^5Q$.  At lowest order, of course, the
renormalization is straightforward: we simply compute the set
of graphs found in Fig.~\ref{fig:fqcd}. The result is finite,
because the current is (partially) conserved, and we extract
a result of the form
\be\label{qcdcurrent}
  \bar q\gamma^\mu\gamma^5 Q\,
  \left(1+\bar\gamma_0{\alpha_s\over4\pi}\,
  \ln(m_Q/m_q)+\dots\right)\,.
\ee
Note that there is no explicit dependence on the
renormalization scale $\mu$, since there is no divergence to
be subtracted.

The same result may be obtained in the effective theory.  In
this case we must match the currents in full QCD onto HQET
currents of the form $\bar q\gamma^\mu\gamma^5 h_v$.  This
step will induce a matching coefficient containing the
explicit dependence on $m_Q$, which is absent, by
construction, from the operators and Lagrangian of the
HQET.\footnote{In general, the matching procedure at order
$\alpha_s$ can also induce new Dirac structures $\bar
q\,\Gamma\, h_v$.  They do not affect the leading logarithms
discussed here.}  In addition, the effective current will
not necessarily be conserved, since the ultraviolet
properties of QCD and the HQET differ.  Hence the form of the
matching, once radiative corrections are included, is
\be
  \bar q\gamma^\mu\gamma^5 Q\to C(m_Q,\mu)\times
  \bar q\gamma^\mu\gamma^5 h_v(\mu)\,.
\ee

\begin{figure}
\epsfysize2.5cm
\hfil\epsfbox{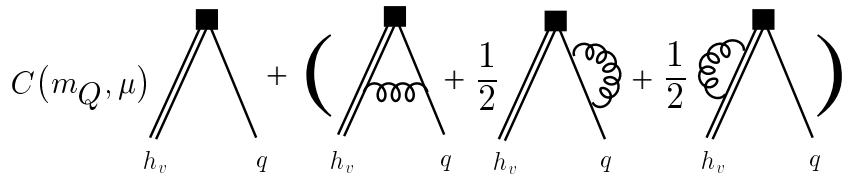}\hfill
\caption{Tree level plus the one loop renormalization of the
current $\bar q\gamma^\mu\gamma^5 h_v$ in HQET.  The double
line is the propagator of the effective field $h_v$.}
\label{fig:fhqet}
\end{figure}

We can deduce the form of $C(m_Q,\mu)$ by considering the
renormalization of the effective current $\bar
q\gamma^\mu\gamma^5 h_v$, shown in the last three terms of
Fig.~\ref{fig:fhqet}.  These diagrams, computed in the
effective theory, are independent of $m_Q$.  However, in
general they are divergent, so they depend on the
renormalization scale $\mu$; the renormalization takes the
form
\be\label{gammadef}
  \bar q\gamma^\mu\gamma^5 h_v(\mu)=
  \bar q\gamma^\mu\gamma^5 h_v(m_q)
  \times\left(1+\gamma_0
  {\alpha_s\over4\pi}\,
  \ln(\mu/m_q)+\dots\right),
\ee
where here $m_q$ acts as an infrared cutoff.  The $\mu$
dependence in the second term must be canceled by
$C(m_Q,\mu)$.  Since the logarithm depends on a
dimensionless ratio, $C(m_Q,\mu)$ must be of the form
\be
  C(m_Q,\mu)=1+\gamma_0{\alpha_s\over4\pi}\,
  \ln(m_Q/\mu)+\dots\,.
\ee
Comparing the dependence on $m_Q$ of $C(m_Q,\mu)$ and the
expansion (\ref{qcdcurrent}), we see that
$\bar\gamma_0=\gamma_0$.

However, the effective theory allows us to go beyond leading
order, and to resum all corrections of the form
$\alpha_s^n\ln^n m_Q$.  We do this with the renormalization
group equations, which express the independence of physical
observables on the renormalization scale $\mu$.  In this
case, they require that the $\mu$ dependence of $C(m_Q,\mu)$
cancel that of the one loop diagrams in Fig.~\ref{fig:fhqet},
under small changes in $\mu$:
\be
  \mu{{\rm d}\over{\rm d}\mu}\,C(m_Q,\mu)
  =-\gamma_0\,{\alpha_s(\mu)\over4\pi}\,.
\ee
The logarithms are resummed because the partial derivative
is promoted to a total derivative with respect to
$\mu$, including the implicit dependence on $\mu$ of the
coupling constant
$\alpha_s(\mu)$:
\bea
  &&\mu{{\rm d}\over{\rm d}\mu}=\mu{\partial\over\partial\mu}
  +\beta(g){\partial\over\partial g}\nonumber\\
  &&\beta(g)=-\beta_0{g^3\over16\pi^2}+\dots\nonumber\\
  &&\beta_0=11-{2\over3}N_f={25\over3}{\rm \ \ for\ }N_f=4\,,
\eea
where $N_f$ is the number of light flavors.  We compute the
anomalous dimension $\gamma_0$ from the ultraviolet divergent
parts of the one loop diagrams shown in Fig.~\ref{fig:fhqet}.

It is instructive to perform the radiative correction to the
current in detail, since this is different from the diagrams
one is used to in ordinary QCD.  With the HQET Feynman rules,
the diagram may be written in Feynman gauge as
\be
  C_f\,(ig)^2\mu^\epsilon\int{{\rm
  d}^{4-\epsilon}q\over(2\pi)^
  {4-\epsilon}}\,
  \overline v_\ell\gamma_\alpha\,{i\over\slash\!\!\!{q}}\,
  \gamma^\mu\gamma^5\,{i\over v\cdot q}\,v^\alpha\,u_h\,
  \times{-i\over q^2}\,,
\ee
where $C_f={4\over3}$ is the color factor.  This expression
may be simplified to
\be
  -C_f\,ig^2\overline
  v_\ell\slash\!\!\!v\gamma_\beta\gamma^\mu
  \gamma^5 \,u_h\,\mu^\epsilon\int{{\rm
  d}^{4-\epsilon}q\over(2\pi)^
  {4-\epsilon}}\,{q^\beta\over q^4v\cdot q}\,,
\ee
which by Lorentz invariance is simply
\be
  -C_f\,ig^2\overline v_\ell\gamma^\mu
  \gamma^5 \,u_h\,\mu^\epsilon\int{{\rm
  d}^{4-\epsilon}q\over(2\pi)^
  {4-\epsilon}}\,{1\over q^4}\,.
\ee
Rotating to Euclidean space, performing the integral and
extracting the pole in $\epsilon$, we find
\be
  \overline v_\ell\gamma^\mu\gamma^5 \,u_h  \times (2C_f)\,
  {g^2\,\mu^\epsilon\over16\pi^2\epsilon}+{\rm finite}\,.
\ee
Since $\mu^\epsilon=1+\epsilon\ln\mu+\dots$, the one loop
contribution to the matrix element then depends on $\ln\mu$ as
\be
  \overline v_\ell\gamma^\mu\gamma^5 \,u_h  \times (2C_f)\,
  {\alpha_s\over4\pi}\ln\mu\,.
\ee
 The contribution to the anomalous dimension is
then
$2C_f={8\over3}$.

The calculation of the wavefunction renormalization of the
heavy quark (the third diagram in Fig.~\ref{fig:fhqet}) is
similar, but requires a novel version of the Feynman trick,
\be
  {1\over ab}=\int^\infty_0 {{\rm d}\lambda\over (a+\lambda
  b)^2}\,.
\ee
One also has to pick out the term which cancels the
$1/v\cdot k$ pole in the heavy quark propagator.  With
the factor of ${1\over2}$ which accompanies the contributions
from wavefunction renormaliaztion, the result is
\be
  {1\over2}\times
  \overline v_\ell\gamma^\mu\gamma^5 \,u_h  \times (4C_f)\,
  {\alpha_s\over4\pi}\ln\mu\,.
\ee
Including the usual QCD renormalization of the light quark
field, we find from the three terms in Fig.~\ref{fig:fhqet},
respectively,\cite{VS,PW}
\be
  \gamma_0={8\over3}+{1\over2}\left(-{8\over3}\right)
  +{1\over2}\left({16\over3}\right)=4\,.
\ee
The solution of the renormalization group equation is
\be
  C(m_Q,\mu)=\left({\alpha_s(m_Q)\over\alpha_s(\mu)}\right)
  ^{-\gamma_0/2\beta_0}
  =\left({\alpha_s(m_Q)\over\alpha_s(\mu)}\right)^{-6/25}\,.
\ee
This then yields the leading logarithmic correction to the
ratio $f_B/f_D$:
\be
  {f_B\over f_D}=\sqrt{{m_D\over m_B}}
  \left({\alpha_s(m_c)\over\alpha_s(m_b)}\right)^{6/25}\,.
\ee
The radiative correction is approximately a ten percent
effect.  In fact, it has a simple physical interpretation. 
For virtual gluons of ``intermediate'' energy,
$m_c<E_g<m_b$, the bottom quark is heavy but the charm quark
is light.  Such gluons contribute to the difference between
$f_B$ and $f_D$ even in the heavy quark limit.

In summary, then, the purpose of the HQET is to make explicit
all dependence of observable quantities on $m_Q$.  The
logarithmic dependence, through $\alpha_s(m_Q)$, arises from
intermediate virtual gluons with $m_c<E_g<m_b$.  We obtain
these corrections by computing perturbatively with the HQET
Lagrangian, then using the renormalization group to resum
the logarithms to all orders.  The power dependence, $1/m_Q$,
is extracted systematically in the heavy quark expansion. 
We have seen how to expand the Lagrangian and the states to
subleading order; the application of the expansion to a
physical decay rate will be presented in the next section.

These lectures are meant to be pedagogical, so we will only
treat the leading corrections to a few processes.  However,
the state of the art goes significantly beyond what will be
presented here.  For many quantities, not only the leading
logarithms, $\alpha_s^n\ln^n m_Q$, but the subleading (two
loop) logarithms, of order $\alpha_s^{n+1}\ln^n m_Q$, have
been resummed.  Similarly, many power corrections are known
to relative order $1/m_Q^2$.  It is particularly important
phenomenologically to take into account the corrections of
order $1/m_c^2$.

\section{Exclusive $B$ Decays}

We now have the tools we need for an HQET treatment of the
exclusive semileptonic transitions $B\to D\,\ell\bar\nu$ and
$B\to D^*\,\ell\bar\nu$.  Earlier, we argued on physical
grounds that in the heavy quark limit all of the hadronic
matrix elements which appear in these decays are related to a
single nonperturbative function $\xi(w)$.  Now we will
sharpen this analysis to actually {\it derive\/} these
relations, and to include radiative and power corrections. 
In fact, almost all of our effort will go into the power
corrections, since the radiative corrections to the
transition currents are computed just as in the previous
section.

\subsection{Matrix element relations at leading order}

The transitions in question require the nonperturbative
matrix elements
\be
  \langle D(v')|\,\bar c\gamma^\mu b\,|B(v)\rangle\,,\quad
  \langle D^*(v',\epsilon)|\,
  \bar c\gamma^\mu b\,|B(v)\rangle\,,\quad
  \langle D^*(v',\epsilon)|\,
  \bar c\gamma^\mu\gamma^5 b\,|B(v)\rangle\,,
\ee
parameterized in terms of form factors as in
Eq.~(\ref{formfactors}).  Our first task is to derive the
relations between these form factors, as promised earlier. 
These relations depend on the heavy quark symmetry, that is,
on the fact that the spin quantum numbers of $Q$ and of the
light degrees of freedom are separately conserved by the soft
physics.  Hence we need a representation of the heavy meson
states in which they have well defined transformations
separately under the angular momentum operators $S_Q$ and
$J_\ell$.  In particular, the representation must reflect
the fact that a rotation by $S_Q$ can exchange the
pseudoscalar meson $M(v)$ with the vector meson
$M^*(v,\epsilon)$.

The solution is to introduce a ``superfield'' ${\cal M}(v)$,
defined as the $4\times4$ Dirac matrix~\cite{FGGW,Falk92}
\be
  {\cal M}(v)={1+\rlap/v\over2}\left[\gamma^\mu 
  M^*_\mu(v,\epsilon)
  -\gamma^5 M(v)\right]\equiv V(v,\epsilon)+P(v)\,.
\ee
Under heavy quark spin rotations $S_Q$, ${\cal M}(v)$
transforms as
\be\label{Mtransform}
  {\cal M}(v)\to D(S_Q){\cal M}(v)\,,
\ee
and under Lorentz rotations $\Lambda$, as
\be
  {\cal M}(v)\to D(\Lambda){\cal M}(\Lambda^{-1}v)D^{-1}
  (\Lambda)\,.
\ee
Here $D(\cdots)$ is the spinor representation of $SO(3,1)$. 
The superfield satisfies the matrix identity
\be\label{Midentity}
  P_+\,{\cal M}(v)\,P_-={\cal M}(v)\,,
\ee
so it transforms the same way as the product of spinors
$h_v\,\bar q$, representing a heavy quark and a light
antiquark moving together at velocity $v^\mu$.

It is straightforward to verify the transformation properties
of the superfield in the rest frame, in which
$v^\mu=(1,0,0,0)$.  In this frame, the spinor representation
of the angular momentum operator ${\bf J}$ has components
$S^i={1\over2}\gamma^5\gamma^0\gamma^i$.  It acts on the
superfield by ${\bf J}^i{\cal M}=[S^i,{\cal M}]$.  Defining
the polarization vectors
$\epsilon^\mu_\pm=(0,1/\sqrt2,\pm i/\sqrt2,0)$ and
$\epsilon^\mu_3=(0,0,0,1)$, it is easy to check that
\be
  {\bf J}^2 P=J^3P=0\,,\quad {\bf J}^2V(\epsilon)=
  2V(\epsilon)\,, \quad {\bf J}^3V(\epsilon_\pm)=\pm
  V(\epsilon_\pm)\,,\quad {\bf J}^3 V(\epsilon)=0\,,
\ee
so $P$ has spin zero and $V(\epsilon)$ spin one.  On the
other hand the heavy quark spin operator ${\bf S_Q}$ has the
same component representation but acts only on the left,
$({\bf S_Q})^i{\cal M}=S^i{\cal M}$.  One may then check that
\be
  ({\bf S_Q})^3P={1\over2}V(\epsilon_3)\,,\qquad
  ({\bf S_Q})^3V(\epsilon_3)={1\over2}P\,.
\ee
As promised, heavy quark spin tranformations exchange the
pseudoscalar and vector mesons.

A current which mediates the decay of one heavy quark ($Q$)
into another ($Q'$) is of the form $\bar
h_{v'}\,\Gamma\,h_v$.  Under a rotation by $S_Q$, the
effective field $h_v$ transforms as
\be
  h_v\to D(S_Q)\,h_v\,,
\ee
while $h_{v'}$ is unchanged.  The current would remain
invariant if we took $\Gamma$ to transform as
\be
  \Gamma\to\Gamma\,D^{-1}(S_Q)\,.
\ee
On the other hand, the matrix element of superfields
\be
  \langle {\cal M}'(v')|\,\bar h_{v'}\,\Gamma\,h_v\,
 |{\cal M}(v)\rangle
\ee
is invariant if we rotate {\it both\/} $h_v$ and ${\cal
M}(v)$ by the {\it same} $S_Q$.  With the transformation law
(\ref{Mtransform}) for ${\cal M}(v)$, it follows that the
$S_Q$-invariant matrix element must be proportional to
$\Gamma{\cal M}(v)$.   When we also consider rotations under
$S_{Q'}$, we find that the matrix element is restricted to
the general form
\be\label{matelgen}
  \langle {\cal M}'(v')|\,\bar h_{v'}\,\Gamma\,h_v\,
  |{\cal M}(v)\rangle=
  -\sqrt{M_MM_{M'}}\,{\rm Tr}\,
  \left[\overline{\cal M}'(v')\,\Gamma\,{\cal M}(v)\,
  \hat F(v,v')\right].
\ee
The product of masses in front is a convention which restores
the relativistic normalization of the states.  Note that the
heavy quark symmetry allows an arbitrary $4\times4$ Dirac
matrix $\hat F(v,v')$ to act on the ``light quark'' part of
the superfields.  Its presence reflects the fact that only Lorentz
symmetry constrains the spin of the light degrees of freedom
during the decay.

A general expansion of $\hat F(v,v')$ in terms of scalar
functions $F_i(w=v\cdot v')$ takes the form
\be
  \hat F(v,v')=F_1(w)+F_2(w)\rlap/v+F_3(w)\rlap/v'+
  F_4(w)\rlap/v\rlap/v'\,.
\ee
However, the identity (\ref{Midentity}) applied to the matrix
element (\ref{matelgen}) yields
\be
  \hat F(v,v')=P_-\,\hat F(v,v')\,P'_-=\left[F_1(w)-
  F_2(w)-F_3(w)+F_4(w)
  \right]P_-\,P'_-\,.
\ee
In other words, $\hat F(v,v')$ actually may be taken to be a
scalar, which we identify with the Isgur-Wise function,
\be
   \hat F(v,v')=\xi(w)\,.
\ee
As an exercise, let us apply this formalism to the matrix
elements for $B\to(D,D^*)\,\ell\bar\nu$.  For a given matrix
element, we pick out the part of the superfield ${\cal M}(v)$
which is relevant.  Hence we find
\bea
  \langle D(v')|\,\bar c\gamma^\mu b\,|B(v)\rangle &=&
  \langle M_D(v')|\,\bar h^c_{v'}\gamma^\mu h^b_v\,|M_B(v
  \rangle\nonumber\\
  &=& -\sqrt{m_Dm_B}\,{\rm Tr}\,
  \left[\gamma^5P'_+\gamma^\mu P_+(-\gamma^5)\right]\,\xi(w)
  \nonumber\\
  &=&\sqrt{m_Dm_B}\,\xi(w)\,(v+v')^\mu\\ \nonumber
  \\
 \langle D^*(v',\epsilon)|\,\bar c\gamma^\mu\gamma^5 b\,|B(v)
  \rangle &=&
  \langle M^*_D(v',\epsilon)|\,\bar h^c_{v'}\gamma^\mu\gamma^5
  h^b_v\,
  |M_B(v)\rangle\nonumber\\
  &=&   -\sqrt{m_{D^*}m_B}\,{\rm Tr}\,
  \left[\rlap/\epsilon P'_+\gamma^\mu\gamma^5
   P_+(-\gamma^5)\right]\,\xi(w)\\
  &=&\sqrt{m_{D^*}m_B}\,\xi(w)\,\left[(w+1)
  \epsilon^{*\mu}
  -\epsilon^*\cdot(v+v')^\mu\right]\nonumber\\ \nonumber\\
  \langle D^*(v',\epsilon)|\,\bar c\gamma^\mu b\,|B(v)
  \rangle &=&
  \langle M^*_D(v',\epsilon)|\,\bar h^c_{v'}\gamma^\mu 
  h^b_v\,
  |M_B(v)\rangle\nonumber\\
  &=&   -\sqrt{m_{D^*}m_B}\,{\rm Tr}\,
  \left[\rlap/\epsilon P'_+\gamma^\mu P_+(-\gamma^5)\right]
  \,\xi(w)\nonumber\\
  &=&\sqrt{m_{D^*}m_B}\,\xi(w)\,
  i\varepsilon^{\mu\nu\alpha\beta}\epsilon_\nu^*v'_\alpha 
  v_\beta\,,
\eea
reproducing explicitly the relations (\ref{ffrelations})
between the independent form factors $h_i(w)$.  We can also
derive the normalization condition at $w=1$.  Consider the
matrix element of the $b$ number current $\bar b\gamma^\mu b$
between $B$ meson states.  In QCD, the matrix element of this
current is exactly normalized,
\be
  \langle B(v)|\,\bar b\gamma^\mu b\,|B(v)\rangle=
  2p_B^\mu=2m_B v^\mu\,.
\ee
But in HQET, we have
\bea
  \langle B(v)|\,\bar b\gamma^\mu b\,|B(v)\rangle&=&
  \langle M_B(v)|\,\bar h_v\gamma^\mu h_v\,|M_B(v)\rangle
  \nonumber\\
  &=& m_B\xi(v\cdot v)(v+v)^\mu\nonumber\\
  &=&2m_Bv^\mu\,\xi(1)\,.
\eea
Hence the normalization condition at zero recoil,
\be
  \xi(1)=1\,,
\ee
follows directly from the conservation of the heavy quark
number current.

\subsection{Power corrections to the matrix elements}

The matrix elements we have derived are computed in the
strict limit $m_{b,c}\to\infty$.  How are they affected by
corrections of order $1/m_b$ and $1/m_c$?  There are two
sources of $1/m_Q$ corrections in the effective theory: the
corrections (\ref{current}) to the heavy quark currents, and
the corrections (\ref{hqetlagrangian}) to the Lagrangian.

When radiative corrections are included, the expansion of
the heavy quark current $\bar c\Gamma b$  in terms of HQET
operators has a form which is somewhat more general than
Eq.~(\ref{current}),
\be
  \bar c\Gamma b\to a_0(\alpha_s)\,\bar h^c_{v'}\Gamma h^b_v
  +{a_1(\alpha_s)\over2m_b}\,\bar h^c_{v'}\Gamma_1^\alpha 
  iD_\alpha h^b_v
  +{a_1'(\alpha_s)\over2m_c}\,\bar h^c_{v'}
  (-i\overleftarrow D_\alpha){\Gamma_1'}^\alpha h^b_v
  +\dots\,.
\ee
The matrix elements of the power corrections are constrained
by heavy quark symmetry in a manner completely analogous to
the leading current.  In terms of traces over the
superfields, we have~\cite{Luke}
\be
  \langle {\cal M}'(v')|\,\bar h_{v'}\,\Gamma^\alpha 
  iD_\alpha\,h_v\,
  |{\cal M}(v)\rangle=
  -\sqrt{M_MM_{M'}}\,{\rm Tr}
  \left[\overline{\cal M}'(v')\,\Gamma^\alpha\,{\cal M}(v)\,
  \hat G_\alpha(v,v')\right],
\ee
where $\hat G_\alpha(v,v')$ is another arbitrary $4\times4$
Dirac matrix.  The matrix element 
\be
  \langle {\cal M}'(v')|\,\bar h_{v'}\,
  (-i\overleftarrow D_\alpha){\Gamma_1'}^\alpha h_v\,|
  {\cal M}(v)\rangle
\ee
may also be written in terms of $\hat G_\alpha(v,v')$, using
charge conjugation.

The $1/m_Q$ corrections ${\cal O}_K$ and ${\cal O}_G$ to the
Lagrangian contribute somewhat differently.   In order to
apply heavy quark symmetry, the matrix elements of the local
currents, both leading and subleading, must be written in
terms of the {\it effective\/} states $|M(v)\rangle$. 
However, these states are not eigenstates of the Hamiltonian,
once ${\cal O}_K$ and ${\cal O}_G$ are included in the
Lagrangian.  Hence, for example, we must allow for the possibility
that if an effective state $|M(v)\rangle$ is created at time
$t=-\infty$, then ${\cal O}_K$ or ${\cal O}_G$ could act on
the state before its decay at $t=0$.  This possibility is
accounted for by including time-ordered products in which
${\cal O}_K$ or ${\cal O}_G$ is inserted along the incoming
or outgoing heavy quark line.  If we are keeping terms of
order $1/m_Q$, only one insertion of ${\cal O}_K$ or ${\cal
O}_G$ needs to be included.  The time-ordered products are of
the form~\cite{Luke}
\bea
  \langle D^{(*)}(v)|&\bar c\Gamma b&|B(v)\rangle =
  \nonumber\\
  \dots &+&
  {1\over2m_c}\,\langle {\cal M}'(v')|\,i\int{\rm d}y\,
  T\left\{
  \bar h^c_{v'}\Gamma h^b_v,{\cal O}^{v'}_K+{\cal O}^{v'}_G
  \right\}
  |{\cal M}(v)\rangle\nonumber\\
  &+&{1\over2m_b}\,\langle {\cal M}'(v')|\,i\int{\rm d}y\,T
  \left\{
  \bar h^c_{v'}\Gamma h^b_v,{\cal O}^v_K+{\cal O}^v_G\right\}
  |{\cal M}(v)\rangle,
\eea
where the ellipses include the current corrections computed
earlier.  The evaluation of the matrix elements of the
time-ordered products will lead to still more
nonperturbative functions like $\hat F(v,v')$ and $\hat
G_\alpha(v,v')$.

\subsection{Corrections at zero recoil}

It is straightforward, but not very illuminating, to expand
all of the new nonperturbative functions which arise at order
$1/m_Q$ in terms of scalar form factors.  In the end, the
corrections may be parameterized in terms of four functions
of the velocity transfer $w$, and a single nonperturbative
parameter $\bar\Lambda$, all proportional to the mass scale
$\lqcd$.\cite{Luke}  The new parameter has a simple
interpretation as the ``energy'' of the light degrees of freedom,
and is given by
\be\label{lambardef}
  \bar\Lambda=\lim_{m_b\to\infty}\,(m_B-m_b)\,.
\ee

Instead of a general treatment, however, we will consider the
$1/m_Q$ corrections at the zero recoil point $w=1$.  This is
clearly the most important case, because it is at this point
that the nonperturbative matrix elements are absolutely
normalized in the heavy quark limit.  What happens to this
normalization condition when $1/m_Q$ corrections are
included?

Let us study the corrections to the current in detail.  They
are described by the nonperturbative function $\hat
G_\alpha(v,v')$.  At $v=v'$, $\hat G_\alpha(v,v)$ may be
expanded as
\be
  \hat G_\alpha(v,v)=G_1 v_\alpha+G_2 \gamma_\alpha
  +G_3 v_\alpha\rlap/v
  +G_4 \gamma_\alpha\rlap/v\,.
\ee
But $\hat G_\alpha(v,v)$ is subject to the same constraint as
$\hat F(v,v')$,
\be
  \hat G_\alpha(v,v)=P_-\hat G_\alpha(v,v)P_-
  =(G_1-G_2-G_3+G_4)v_\alpha P_-
  \equiv G\,v_\alpha P_-\,,
\ee
and it, too, is equivalent to a Dirac scalar (the same is
{\it not\/} true of the general function $\hat
G_\alpha(v,v')$).  Now consider the matrix element where we
take $\Gamma^\alpha=v^\alpha$.  Then we have
\bea
  \langle {\cal M}'(v)|\,\bar h_{v'}\,iv\cdot D\,h_v\,
  |{\cal M}(v)\rangle
  &=&\mbox{}-\sqrt{M_MM_{M'}}\,{\rm Tr}\,
  \left[\overline{\cal M}'(v)\,v^\alpha\,{\cal M}(v)\right]\,
  G\,v_\alpha\nonumber\\
  &=& \mbox{}-G\sqrt{M_MM_{M'}}\,{\rm Tr}\,
  \left[\overline{\cal M}'(v)\,{\cal M}(v)\right].
\eea
But this matrix element {\it vanishes\/} by the classical
equation of motion in the effective theory,
\be
  v\cdot D\,h_v(x)=0\,.
\ee
Hence $G=0=\hat G_\alpha(v,v)$.  There are no $1/m_Q$
corrections from the current to the normalization condition
at zero recoil.\cite{Luke,CG}

The same is true of insertions of the corrections ${\cal
O}_K$ and ${\cal O}_G$ to the Lagrangian: their contribution
vanishes at $w=1$.  To show this requires the imposition of
the conservation of the $b$ number current at order $1/m_b$,
much as we derived the normalization of the Isgur-Wise
function at leading order.  This part of the argument is
analogous to the classic nonrenormalization theorem of
Ademollo and Gatto.\cite{AdGa64}

In the end, we have the result known as
{\it Luke's Theorem}.\cite{Luke}  There are no corrections
at zero recoil to the hadronic matrix elements responsible
for the semileptonic decays $B\to D\,\ell\bar\nu$ and $B\to
D^*\,\ell\bar\nu$.  The leading power corrections to the
normalization of zero recoil matrix elements are only of
order $1/m_c^2$.  Given that $\lqcd/m_c\sim30\%$ and
$\lqcd^2/m_c^2\sim10\%$, the implication is that the leading
order predictions at $w=1$ are considerably more accurate
than one might have expected.  In addition, away from zero
recoil the $1/m_c$ corrections must be suppressed at least by
$(w-1)$.

On closer inspection, this result is more interesting for
$B\to D^*\,\ell\bar\nu$ than for $B\to D\,\ell\bar\nu$. 
This is because the leading order matrix element for $B\to
D\,\ell\nu$ vanishes kinematically at zero recoil for a
massless lepton in the final state.  Hence, in this case the
$1/m_c$ corrections are not suppressed as a fractional
correction to the lowest order term.\cite{NR}

\subsection{Extraction of $|V_{cb}|$ from $B\to
D^*\,\ell\bar\nu$}

An immediate application of these results is the extraction
of $|V_{cb}|$ from the exclusive decay $B\to
D^*\,\ell\bar\nu$.  This process is mediated by the weak
operator ${\cal O}_{bc}$~(\ref{Obcdef}), whose matrix
element factorizes as
\be
  \langle D^*\,\ell\bar\nu|\,{\cal O}_{bc}\,|B\rangle=
  {G_FV_{cb}\over\sqrt2}\,
  \langle D^*|\,\bar c\gamma^\mu(1-\gamma^5)b\,|B\rangle\,
  \langle\ell\bar\nu|\,\bar\ell\gamma_\mu(1-\gamma^5)\nu\,
  |0\rangle\,.
\ee
The leptonic matrix element may be computed perturbatively,
while we treat the  hadronic matrix element in the heavy
quark expansion.  The result is a differential decay rate of
the form~\cite{Neu91}
\bea\label{diffrate}
  {{\rm d}\Gamma\over{\rm d}w}&=&{G_F^2\over48\pi^3}\,
  |V_{cb}|^2\,
  (m_B-m_{D^*})^2m_{D^*}^3(w+1)^3\sqrt{w^2-1}\nonumber\\
  &&\quad\times\left[1+{4w\over w+1}\,{m_B^2-2wm_bm_{D^*}
  +m_{D^*}^2
  \over(m_B-m_{D^*})^2}\right]F^2(w)\,.
\eea
All of the HQET analysis goes into the factor $F(w)$, which
has an expansion
\be
  F(w)=\xi(w)+{\rm (radiative\ corrections)}+
  {\rm (power\ corrections)}\,.
\ee

We extract $|V_{cb}|$ by studying the differential decay rate
near $w=1$, where the hadronic matrix elements are known.  Of
course, this requires extrapolation of the experimental data,
since the rate vanishes kinematically at $w=1$.  For massless
leptons, only the matrix element $\langle D^*|\,\bar
c\gamma^\mu\gamma^5b\,|B\rangle$ of the axial current
contributes at this point.  The analysis of this quantity in
the HQET yields an expansion of the form
\be
  F(1) =\eta_A\,\left[1+{0\over m_c}+{0\over m_b}
  +\delta_{1/m^2}+\dots\right]\,.
\ee
The correction $\delta_{1/m^2}$, which contains terms
proportional to $1/m_c^2$, $1/m_b^2$ and $1/m_cm_b$, is
intrinsically nonperturbative.  It has been estimated from a
variety of models to be small and
negative,\cite{FN,SUV,Neu94}
\be
  \delta_{1/m^2}\approx -0.055\pm0.035\,.
\ee
Note that the model dependence in the result has been
relegated to the estimation of the sub-subleading terms.  The
radiative correction $\eta_A$ has now been computed to two
loops,\cite{Cz,FGN}
\be
  \eta_A=0.960\pm0.007\,.
\ee
The result is a value for $F(1)$ with errors at the level of
5\%,
\be
  F(1)=0.91\pm0.04\,.
\ee
This is the theory error which the experimental determination
of $|V_{cb}|$ will inherit.  It is dominated by the
uncertainty in the nonperturbative corrections, and it is
difficult to see how this can be improved much in the future.

All that is left experimentally is to extrapolate the data to
$w=1$ and extract
\be
  \lim_{w\to1}\,{1\over\sqrt{w-1}}\,
  {{\rm d}\Gamma\over{\rm d}w}\,.
\ee
Once the kinematic factors in Eq.~(\ref{diffrate}) have been
included, this amounts to a direct measurement of the
combination $|V_{cb}|F(1)$.  Both CLEO and LEP have
reported results for this quantity.\cite{CL1,LHFWG}  They
have taken the slightly different value $F(1)=0.88\pm0.05$,
and I have scaled up the theory error of the CLEO result to
make it consistent with LEP.  Then we have quite consistent
results for $V_{cb}$:
\bea
  {\rm CLEO:}&\qquad&(39.4\pm2.1\pm2.0\pm2.2)\times10^{-3}
  \nonumber\\
  {\rm LEP\ average:}
&\qquad&(38.4\pm1.1\pm2.2\pm2.2)\times10^{-3}\,.
\eea
This value of $|V_{cb}|$ has almost no dependence on hadronic
models.  In contrast to model-based ``measurements'', here
the theoretical error is meaningful, in that it is based on a
systematic expansion in small quantities.

\section{Inclusive $B$ Decays}

An exclusive semileptonic $B$ decay, such as $B\to
D\,\ell\bar\nu$, is one in which the final hadronic state is
fully reconstructed.  An inclusive decay, by contrast, is one
in which only certain kinematic features, and perhaps the
flavor, of the hadron are known.  In this case, we need a
theoretical analysis in which we sum over all possible
hadronic final states allowed by the kinematics. 
Fortunately, this is possible within the structure of the
HQET.

As in the case of exclusive decays, the key theme is the
separation of short distance physics, associated with the
heavy quark, from long distance physics, associated with the
light degrees of freedom.  We will also rely on heavy quark
spin and flavor symmetry.  However, the new ingredient will
be the idea of ``parton-hadron duality'', which, as we will
see, also relies on the heavy quark limit $m_b\gg\lqcd$.

\subsection{The inclusive decay $B\to X_c\,\ell\bar\nu$}

Let us consider the inclusive decay
\be
  B(p_B)\to X_c(p_X)\,\ell(p_\ell)\nu(p_\nu)\,,
\ee
where all that is known about the state $X_c$ are its energy
and momentum, and the fact that it contains a charm quark. 
This decay is mediated by the weak operator ${\cal O}_{bc}$. 
It is easy to generalize our discussion to inclusive decays
of other heavy quarks, such as $b\to u\,\ell\nu$ and $c\to
s\,\bar\ell\nu$, by replacing ${\cal O}_{bc}$ with the
appropriate weak operator.

The treatment of exclusive decays required both the $b$ and
$c$ quarks to be heavy.  For inclusive decays we can relax
this condition on the $c$ quark, requiring only
$m_b\gg\lqcd$.  What does the weak decay of the $b$, at time
$t=0$, look like to the light degrees of freedom?  For $t<0$,
there is a heavy hadron composed of a point-like color source
and light quarks and gluons.  At $t=0$, the point source
disappears, releasing both its color and a large amount of
energy into the hadronic environment.  Eventually, for $t>0$,
this new collection of strongly interacting particles will
materialize as a set of physical hadrons.  The probability of
this hadronization is unity; there is no interference between
the hadronization process and the heavy quark decay.  There
are subleading effects in powers of $\lqcd/m_b$, but they do
not alter the probability of hadronization.  Rather, they
reflect the fact that the $b$ quark is not exactly a static
source of color:  it has a small nonrelativistic kinetic
energy and it carries a spin, both of which affect the
kinematic properties of its decay.

As in the case of exclusive decays, we will compute the
inclusive semileptonic width $\Gamma(B\to X_c\,\ell\bar\nu)$
as a double expansion in powers of $\alpha_s(m_b)$ and
$\lqcd/m_b$.\cite{FLS94,CGG,SV,MW}  The expansion in
$\alpha_s(m_b)$ reflects the applicability of perturbative
QCD to the short distance part of the process.  The heavy
quark expansion will be continued to relative order
$1/m_b^2$, as there is an analogue of Luke's Theorem which
eliminates power corrections to the rate of order $1/m_b$. 
These corrections will be written in terms of three
nonperturbative parameters.  The first, $\bar\Lambda$, is
defined in Eq.~(\ref{lambardef}).  It is essentially the mass
of the light degrees of freedom  in the heavy hadron, but we
will see that it is plagued by an ambiguity of order $\lqcd$
in the definition of the $b$ quark mass.  The other two
parameters are the expectation values in the $B$ meson of the
leading corrections ${\cal O}_K$ and ${\cal O}_G$ to
$\lhqet$.  They are defined as~\cite{FN}
\bea
  \lambda_1&=&{1\over2m_B}\langle B|\,{\cal O}_K\,
  |B\rangle\nonumber\\
  \lambda_2&=&-{1\over6m_B}\langle B|\,{\cal O}_G\,
  |B\rangle\,,
\eea
where we take the usual relativistic normalization of the
states.  Hence, $\lambda_1$ may be thought of roughly as the
negative of the $b$ quark kinetic energy, and $\lambda_2$ as
the energy of its hyperfine interaction with the light
degrees of freedom.

Now let us outline the computation.  The inclusive decay
involves a sum over all possible final states, which is
actually a sum over exclusive modes (such as $D,
D^*,D\pi,\ldots$), followed by a phase space integral for
each mode.  We write
\be
  \Gamma(B\to X_c\,\ell\bar\nu)=\sum_{X_c}\int{\rm d}
  [{\rm P.S.}]\;
  \big|\langle X_c\,\ell\bar\nu|\,{\cal O}_{bc}\,
  |B\rangle\big|^2\,.
\ee
There is an Optical Theorem for QCD, which follows from the
analyticity of the scattering matrix as a function of the
momenta of the asymptotic states.  Its content is that a
transition rate is proportional to the imaginary part of the
forward scattering amplitude with two insertions of the
transition operator,
\be
  \Gamma(B\to X_c\,\ell\bar\nu)=-2\,{\rm Im}\;i\int{\rm d}x
  \,e^{ik\cdot x}\,
  \langle B|\,T\left\{{\cal O}^\dagger_{bc}(x),
  {\cal O}_{bc}(0)\right\}
  \,|B\rangle\equiv 2\,{\rm Im}\,T\,.
\ee
In what follows, we will write the time-ordered product
$T\{{\cal O}^\dagger_{bc},{\cal O}_{bc}\}$ as a series of
local operators, using the Operator Product Expansion.  As we
will see, the applicability of this expansion, and its
computation in perturbation theory, will rest on the limit
$m_b\gg\lqcd$.  We will then use this limit again to expand
the matrix elements of these local operators in the HQET.

The first step is to factorize the integration over the
lepton momenta, which can be performed explicitly.  Written
as a product of currents, ${\cal O}_{bc}$ takes the form
\be
  {\cal O}_{bc}={G_FV_{cb}\over\sqrt2}\,J^\mu_{bc}\,
  J_{\ell\mu}\,,
\ee
where
\bea
  J^\mu_{bc}&=&\bar c\gamma^\mu(1-\gamma^5)b\nonumber\\
  J^\mu_\ell&=&\bar\ell\gamma^\mu(1-\gamma^5)\nu\,.
\eea
Then $T$ can be decomposed as an integral over the total
momentum $q^\mu=p_\ell^\mu+p_\nu^\mu$ transferred to the
leptons,
\be\label{Tdef}
  T={1\over2}G_F^2|V_{cb}|^2\int{\rm d}q\,T^{\mu\nu}(q)
  \,L_{\mu\nu}(q)\,.
\ee
Here the lepton tensor is
\bea
  L_{\mu\nu}(q)&=&\int{\rm d}[{\rm P.S.}]\;
  \langle0|\,J^\dagger_{\ell\mu}\,|\ell\bar\nu\rangle\,
  \langle\ell\bar\nu|\,J_{\ell\nu}\,|0\rangle\nonumber\\
  &=&{1\over3\pi}\,\left(q_\mu q_\nu-q^2g_{\mu\nu}\right)\,,
\eea
and the hadron tensor is
\be
  T^{\mu\nu}(q)=-i\int{\rm d}x\,e^{iq\cdot x}\,\langle B|
  \,T\left\{
  J^{\mu\dagger}_{bc}(x),J^\mu_{bc}(0)\right\}\,|B\rangle\,.
\ee
We will need the imaginary part, ${\rm Im}\;T^{\mu\nu}$. 
Where is it nonvanishing?  In quantum field theory, a
scattering amplitude develops an imaginary part when there
can be a real intermediate state, that is, the intermediate
particles can all go on their mass shell.  Whether this is
possible, of course, depends on the kinematics of the
external states.  

\begin{figure}
\epsfysize5cm
\hfil\epsfbox{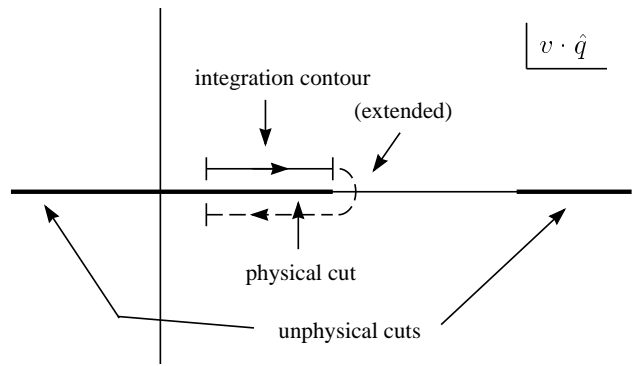}\hfill
\caption{Analytic structure of $T^{\mu\nu}$ in the complex
$v\cdot\hat q$ plane, for fixed, real $\hat q^2$.  The
integration contour is over the ``physical cut'',
corresponding to real decay into leptons.  The unphysical
cuts correspond to other processes.}
\label{fig:vqplane}
\end{figure}

In this case, there are two avenues for creating a physical
intermediate state.\cite{CGG}  The first is to act on the
external state $|B\rangle$ with the transition current
$J^\mu_{bc}$.  The state which is created has no net $b$
number and a single charm quark; the simplest possibility is
the decay process $b\to c$.  The momentum of the intermediate
state is $p_X=p_B-q$; the condition that it could be on mass
shell is simply
\be
  p_X^2=(p_B-q)^2\ge m_D^2\,.
\ee
If we define scaled variables
\be
  p_B^\mu=m_B v^\mu\,,\qquad\hat q^\mu=q^\mu/m_B\,,\qquad
  \hat m_D=m_D/m_B\,,
\ee
this condition becomes
\be
  v\cdot\hat q\le{1\over2}\left(1+\hat q^2-\hat m_D^2\right)\,.
\ee
Another possibility is to act on $|B\rangle$ with the
conjugate operator $J^{\mu\dagger}_{bc}$.  This operation
would produce an intermediate state with two $b$ quarks and
one $\bar c$.  For this state to be on shell, the momentum
transfer has to satisfy
\be
  p_X^2=(p_B+q)^2\ge(2m_B+m_D)^2\,,
\ee
that is,
\be
  v\cdot\hat q\ge{1\over2}\left(3-\hat q^2+4\hat m_D+
  \hat m_D^2\right)\,.
\ee
The physical intermediate states are shown as cuts in the
$v\cdot\hat q$ plane in Fig.~\ref{fig:vqplane}.  Also shown
is the contour corresponding to the  phase space integration
over the lepton momentum $q$.  For physical (massless)
leptons which are the product of a heavy quark decay, this
integral runs over the top of the lower cut, for the range
\be
  \sqrt{\hat q^2}+i\epsilon\le v\cdot\hat q\le
  {1\over2}\left(1+\hat q^2-\hat m_D^2\right)+i\epsilon\,.
\ee
As indicated by the dotted line, we can continue this contour
around the end of the cut and back along the bottom, to
$v\cdot\hat q=\sqrt{\hat q^2}-i\epsilon$.  Since
$T^{\mu\nu}(v\cdot\hat q\,^*)=-T^{\mu\nu}(v\cdot\hat q)$ for
real $\hat q^2$, we compensate for extending the contour by
dividing the new integral by two.

We now encounter our central problem.  The integral over
$v\cdot\hat q$ runs over physical intermediate hadron states,
which are color neutral bound states of quarks and gluons. 
Hence the integrand depends intimately on the details of QCD
at long distances, which is intrinsically nonperturbative.  A
perturbative calculation of $T^{\mu\nu}$, which is all we
have at our disposal, would appear to be of no use.

\begin{figure}
\epsfysize3cm
\hfil\epsfbox{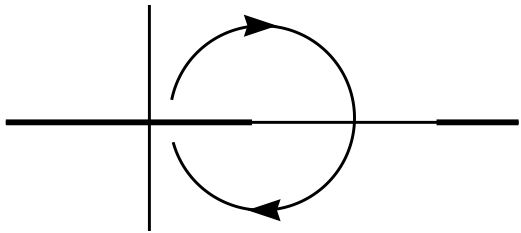}\hfill
\caption{The deformation of the integration contour into the
complex $v\cdot\hat q$ plane.}
\label{fig:contour}
\end{figure}

The solution is to deform the contour away from the cut, into
the complex $v\cdot\hat q$ plane, as shown in
Fig.~\ref{fig:contour}.  Since the scale of momenta is set by
$m_b$, the contour is now a distance of order $m_b$ away from the
resonances.\cite{CGG}  Since $m_b\gg\lqcd$, it is
reasonable to hope that a perturbative treatment in this
region is valid.  Essentially, we are saved because we do not
need to know $T^{\mu\nu}(q)$ for every value of $q$, just
suitable integrals of $T^{\mu\nu}$.  That we can use such
arguments to compute perturbatively the {\it average\/}
value of a hadronic quantity, where at each point the
quantity depends on nonperturbative physics, is known as
(global) {\it parton-hadron duality.}  

Parton-hadron duality has the status of being somewhat more
than an assumption, since it is known to hold in QCD in the
limit $m_b\to\infty$, but somewhat less than an
approximation, since it is not known how to compute
systematically the leading corrections to it.  In any case,
the limit $m_b\gg\lqcd$ plays a crucial role here.  By
deforming the integration contour a distance of order $m_b$ away
from the resonance regime, we find the correspondence in QCD
of our earlier intuitive statement: the
probability of the decay products materializing as physical
hadrons is unity, independent of the kinematics of the short
distance process.  The local redistribution of probability in
phase space due to the presence of hadronic resonances is
irrelevant to the total decay.  Finally, we should note that
since we do not have control over the corrections to local
duality, it might work better in some processes than in
others, for reasons that need not be apparent from within the
calculation.  Hence one must be particularly wary of drawing
dramatic conclusions from any surprising results of these
inclusive calculations.\cite{FDW}

\begin{figure}
\epsfysize3cm
\hfil\epsfbox{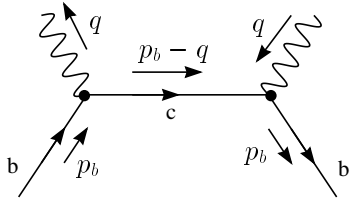}\hfill
\caption{The operator product expansion at tree level.}
\label{fig:top}
\end{figure}

Let us perform the operator product expansion at tree level,
and for decay kinematics.  The Feynman diagram is given in
Fig.~\ref{fig:top}, which yields the expression
\be
  T^{\mu\nu}=
  \bar b\gamma^\mu(1-\gamma^5)\,{\rlap/p_b-\rlap/q+m_c\over
  (p_b-q)^2-m_c^2+i\epsilon}\,\gamma^\nu(1-\gamma^5)b\,.
\ee
We now write
\bea
  p_b^\mu&=&m_bv^\mu +k^\mu=m_b(v^\mu+\hat k^\mu)\nonumber\\
  \hat q^\mu&=&q^\mu/m_b\nonumber\\
  \hat m_c&=&m_c/m_b\nonumber\\
  b(x)&=&e^{-im_bv\cdot x}\,h_v(x)+O(1/m_b)\,,
\eea
and expand in powers of $1/m_b$.  Since the operator product
expansion is in terms of the effective field $h_v$, a factor
of $k^\mu$ corresponds to an insertion of the covariant
derivative $iD^\mu$.  Operator ordering ambiguities are to be
resolved by considering graphs with external gluon fields.

As an example of the procedure, let us expand the propagator
to order $1/m_b^2$.  (There are also corrections to the
currents at this order, which are included in a full
calculation.)  It is convenient to define the scaled hadronic
invariant mass,
\be
  \hat s=(m_bv^\mu-q^\mu)^2/m_b^2=1-2v\cdot\hat q+\hat q^2\,.
\ee
Then we find a contribution to $T^{\mu\nu}$ of the form
\be\label{Texpand}
  T^{\mu\nu}={1\over m_b}\,\bar h_v\gamma^\mu(\rlap/v-\rlap/
  {\hat q})
  \gamma^\nu(1-\gamma^5)\left[{1\over\hat s-\hat m_c^2
  +i\epsilon}
  +{2\hat k\cdot\hat q-\hat k^2\over
  (\hat s-\hat m_c^2+i\epsilon)^2}+\dots\;\right]h_v\,.
\ee
From this expression we can read off the operators which
appear in the operator product expansion.  Since
\be
  {\rm Im}\;{1\over(\hat s-\hat m_c^2+i\epsilon)^n}=
  \pi(n-1)!(-1)^n\delta^{(n-1)}(\hat s-\hat m_c^2)\,,
\ee
we see that the effect of taking the imaginary part in each
term is to put the charm quark on its mass shell.  The
leading term is a quark bilinear,
\be
  {1\over m_b}\,\bar h_v\gamma^\mu(\rlap/v-\rlap/{\hat q})
  \gamma^\nu(1-\gamma^5)h_v\,.
\ee
It is straightforward to compute its matrix element in the
HQET using the trace formalism,
\be
  \langle B|\bar h_v\gamma^\mu(\rlap/v-\rlap/{\hat q})
  \gamma^\nu(1-\gamma^5)h_v|B\rangle
  =2m_B\left(2v^\mu v^\nu-g^{\mu\nu}-v^\mu\hat q^\nu-v^\nu\hat
  q^\mu+g^{\mu\nu}v\cdot\hat q\right).
\ee
The leading corrections to this expression are of order
$1/m_b^2$; there are no corrections of order $1/m_b$, by
Luke's Theorem.  Finally, we contract the tensor $T^{\mu\nu}$
with $L_{\mu\nu}$ and perform the phase space integration
(\ref{Tdef}).  In the end, the result is the same as we would
have gotten directly by computing free quark decay,
\be
  \Gamma={G_F^2|V_{cb}|^2m_b^5\over192\pi^3}\left(
  1-8\hat m_c^2+8\hat m_c^6-\hat m_c^8-12\hat m_c^4
  \ln\hat m_c^2\right).
\ee

Of course, if we only intended to reproduce the free quark
decay result, we would never have introduced so much new
formalism.  The value of the HQET framework is that it allows
us to go beyond leading order and compute the next terms in
the series in $1/m_b^n$.  For example, consider the operators
induced by the expansion of the propagator (\ref{Texpand}). 
The correction of order $1/m_b$ comes from the operator
\be
  {1\over m_b^2}\,{2\over(\hat s-\hat m_c^2+i\epsilon)^2}\,
  \bar h_v\gamma^\mu(\rlap/v-\rlap/{\hat q})
  \gamma^\nu(1-\gamma^5)\hat q\cdot iD\,h_v\,.
\ee
However, the matrix element of this operator is of the form
\be
  \langle B|\,\bar h_v\Gamma^\alpha (v,q)\,iD_\alpha\,h_v\,
  |B\rangle\,,
\ee
which, as we have seen, vanishes by the classical equation of
motion.  (In writing Eq.~(\ref{Texpand}), we have already
dropped terms explicitly proportional to $v\cdot\hat k$,
for the same reason.)  In fact, since all
$1/m_b$ corrections, from any source, have a single covariant
derivative, they all vanish in the same way.  This is the
analogue of Luke's Theorem for inclusive decays.\cite{CGG} 
The correction of order
$1/m_b^2$ in Eq.~(\ref{Texpand}) is
\be
  -{1\over m_b^3}\,{1\over(\hat s-\hat m_c^2+i\epsilon)^2}\,
  \bar h_v\gamma^\mu(\rlap/v-\rlap/{\hat q})
  \gamma^\nu(1-\gamma^5)(iD)^2\,h_v\,.
\ee
The matrix element of this operator is related by the heavy
quark symmetry to $\lambda_1$, the expectation value of
${\cal O}_K$.  The full expansion of $T^{\mu\nu}$ also
induces operators with explicit factors of the gluon field,
whose matrix elements are related to $\lambda_2$.

We now present the result for the inclusive semileptonic
decay rate, up to order $1/m_b^2$ in the heavy quark
expansion, and with the complete radiative correction of
order $\alpha_s$.  We also include that part of the two loop
correction which is proportional to $\beta_0\alpha_s^2$. 
Since $\beta_0\approx9$, perhaps this term dominates the two
loop result.  In any case, it is interesting for other
reasons, as we will see below.

Let us first consider the decay $B\to X_u\,\ell\bar\nu$, for
which the decay rate simplifies since $m_u=0$.  We
find~\cite{SV,MW,GPR,LSW1,FLS96}
\bea\label{burate}
  \Gamma(B\to
  X_u\,\ell\bar\nu)&=&{G_F^2|V_{ub}|^2\over192\pi^3}
  \,m_b^5\,\Bigg[1+\left({25\over6}-{2\pi^2\over3}\right)
  {\alpha_s(m_b)\over\pi}\\ &&
  \mbox{}-(2.98\beta_0+C_u)
  \left({\alpha_s(m_b)\over\pi}\right)^2
  +{\lambda_1-9\lambda_2\over2m_b^2}+\dots\;\Bigg].\nonumber
\eea
When we include the charm mass, it is convenient to write the
unknown quark masses in terms of the measured meson masses
and the parameters of the HQET.  In terms of the spin
averaged mass $\overline m_B=(m_B+3m_{B^*})/4$, we have
\be\label{mexpand}
  m_b=\overline m_B-\bar\Lambda+{\lambda_1\over2m_B}+\dots\,,
\ee
and analogously for $m_c$.  We then find~\cite{SV,MW,Nir,LSW2}
\bea\label{bcrate}
  \Gamma(B\to X_c\,\ell\bar\nu)=&&
  {G_F^2|V_{cb}|^2\over192\pi^3}
  \,m_B^5\times 0.369\,
  \Bigg[1-1.54{\alpha_s(m_b)\over\pi}\nonumber\\ 
  &&\mbox{}-
  (1.43\beta_0+C_c)\left({\alpha_s(m_b)\over\pi}\right)^2
  -1.65{\bar\Lambda\over m_B}
  \left(1-0.87{\alpha_s(m_b)\over\pi}\right)\nonumber\\
  &&\mbox{}-0.95{\bar\lambda^2\over m_B^2}
  -3.18{\lambda_1\over m_B^2}
  +0.02{\lambda_2\over m_B^2}+\dots\;\Bigg].
\eea
All the coefficients which appear in this expression are
known functions of $\overline m_D/\overline m_B$, and are
evaluated at the physical point $\overline m_D/\overline
m_B=0.372$.  In both $B\to X_u\,\ell\bar\nu$ and $B\to
X_c\,\ell\bar\nu$, the power corrections proportional to
$\lambda_1$ and $\lambda_2$ are numerically small, at the
level of a few percent.

\subsection{Renormalons and the pole mass}

The inclusive decay rate depends on the heavy quark mass
$m_b$, either explicitly, as in Eq.~(\ref{burate}), or
implicitly through $\bar\Lambda$, as in Eq.~(\ref{bcrate}). 
At tree level, $m_b$ is just the coefficient of the $\bar
b\,b$ term in the QCD Lagrangian, but beyond that we are
faced with the question of what exactly we mean by $m_b$. 
Should we take an $\overline{\rm MS}$ mass, such as
$\overline m_b(m_b)$?  Or should we take the pole mass
$\mbpole$, or maybe some other quantity?  The various
prescriptions for $m_b$ can vary by hundreds of MeV, and,
since the total rate is proportional to $m_b^5$, the question
is of practical importance if we hope to make accurate
phenomenological predictions.

At a fixed order in QCD perturbation theory, the answer is
clear.  The heavy quark masses which appear come from poles
in quark propagators, so we should take $\mbpole$ (and
$\mcpole$).  This is also the prescription for the mass which
cancels out the on-shell part of the heavy quark field in the
construction of $\lhqet$.  Hence the difference of heavy
quark pole masses is known quite well,
\bea
  \mbpole-\mcpole&=&\left(\overline m_B-\bar\Lambda+
  {\lambda_1\over2m_B}+\dots
  \right)-\left(\overline m_D-\bar\Lambda+
  {\lambda_1\over2m_D}+\dots\right)
  \nonumber\\
  &=&3.34\gev+O(\lqcd^2/m_Q^2)\,.
\eea
Since $\Gamma(B\to X_c\,\ell\bar\nu)$ depends approximately
as $m_b^2(m_b-m_c)^3$, the uncertainty due to quark mass
dependence is reduced.

The problem, of course, is that there is no sensible
nonperturbative definition of $\mbpole$, since due to
confinement there is no actual pole in the quark propagator. 
Hence a direct experimental determination of a value for
$\mbpole$ to insert into the theoretical expressions
(\ref{burate}) and (\ref{bcrate}) is not possible.  How,
then, can we do phenomenology?

\begin{figure}
\epsfysize4cm
\hfil\epsfbox{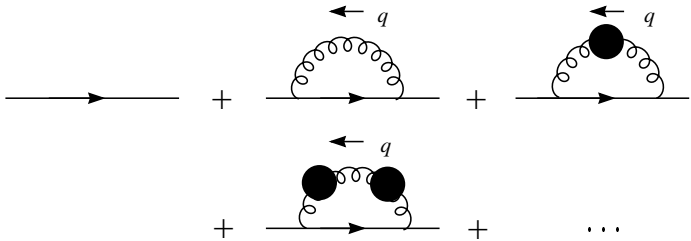}\hfill
\caption{The radiative corrections to $\mbpole$ of order
$\alpha_s(m_b)^{n+1}\beta_0^n$.}
\label{fig:mbubbles}
\end{figure}

One approach would be to define $\mbpole$ to be the pole mass
as computed in perturbation theory, truncate at some order,
and then estimate the theoretical error from the uncomputed
higher order terms.  However, it turns out that even {\it
within\/} perturbation theory the concept of a quark pole
mass is ambiguous.  Consider a particular class of diagrams
which contribute to $\mbpole$, shown in
Fig.~\ref{fig:mbubbles}.  The perturbation theory is
developed as an expansion in the small parameter
$\alpha_s(m_b)$, so we hope that it will be well behaved. 
Each of the bubbles represents an insertion of the gluon
self-energy, which is proportional at lowest order to
$\alpha_s(m_b)\beta_0$.  Of course, the infinite sum of the
graphs in Fig.~\ref{fig:mbubbles} can be absorbed into the
one loop graph, with a compensating change in the coupling
from $\alpha_s(m_b)$ to $\alpha_s(q)$, where $q$ is the loop
momentum.
The result is an expansion for $\mbpole$ of the form
\be
  \mbpole=\overline m_b(m_b)\bigg[1+a_1\alpha_s+
  (a_2\beta_0+b_2)\alpha_s^2
  +(a_3\beta_0^2+b_3\beta_0+c_3)\alpha_s^3+\dots\bigg],
\ee
where $\alpha_s=\alpha_s(m_b)$.  The graphs in
Fig.~\ref{fig:mbubbles} contribute the terms proportional to
$\alpha^{n+1}_s\beta_0^n$.  Since
$\beta_0\approx9$ these terms are ``intrinsically'' larger
than ones with fewer powers of $\beta_0$, and we might hope
that their sum approximates the full series.  However, it is
important to realize that the only limit of QCD in which such
terms actually dominate is that of large number of light
quark flavors, in which case the sign of $\beta_0$ is
opposite to that of QCD.  Although this is a physical limit
of an abelian theory, we are certainly not close to that
limit here.  The ansatz of keeping only the terms
proportional to $\alpha^{n+1}_s(m_b)\beta_0^n$ is known as
``naive nonabelianization'' (NNA).\cite{BBZ}

What is most interesting about the series of terms shown in
Fig.~\ref{fig:mbubbles}, which takes the form $\sum
a_n\alpha^n_s\beta_0^{n-1}$, is that it does not
converge.  Already in the graphs kept in the NNA ansatz, we
are sensitive to the fact that QCD is an asymptotic, rather
than a convergent expansion.  For large $n$ the coefficients
$a_n$ diverge as $n!$, much stronger than any convergence due
to the powers $\alpha^n_s$.  The series can only be made
meaningful if this divergence is subtracted.  As with many
subtraction prescriptions, there is a residual finite
ambiguity.\footnote{In a  formal treatment, this ambiguity
arises from a choice of contour in the Borel plane.}  This
ambiguity, known as an ``infrared renormalon'', leads to an
ambiguity in the pole mass of order~\cite{BBZ,Bigi,NS}
\be
  \delta\mbpole\sim100\mev\,.
\ee
By the definition (\ref{lambardef}), $\bar\Lambda$ also
inherits this ambiguity.

The expressions (\ref{burate}) and (\ref{bcrate}) are plagued
by two problems.  The first is the renormalon ambiguity in
$\mbpole$ and $\bar\Lambda$.  The second is that the
perturbative expansion for the rate $\Gamma$ is itself
divergent, and {\it also\/} has an infrared renormalon.  In
the expansion
\be
  \Gamma=\Gamma_0\left[\sum a'_n\alpha^n_s(m_b)\beta_0^{n-1}+
  {\rm (power\ corrections)\ }\right],
\ee
the coefficients $a_n'$ also diverge as $n!$.  However, it
turns out that these two problems actually cure each other,
because the infrared renormalons in $\mbpole$ and in the
perturbation series for $\Gamma$ cancel.\cite{Bigi,LMS}  We
can exploit this cancelation to improve the predictive power
of the theoretical computation of the rate.  Without this
improvement, the infrared renormalons render the expressions
(\ref{burate}) and (\ref{bcrate}) of dubious
phenomenological utility.

The most reliable approach, theoretically, is to eliminate
$\mbpole$ or $\bar\Lambda$ explicitly from the rate by
computing and measuring another quantity which also depends
on it.  For example, let us consider the charmless decay rate
$\Gamma(B\to X_u\,\ell\bar\nu)$ and the average invariant
mass $\langle s_H\rangle$ of the hadrons produced in the
decay.  Each of these expressions suffers from a poorly
behaved perturbation series in the NNA approximation. 
Ignoring terms of relative order $1/m_b^2$ and writing the
rate in terms of $\bar\Lambda$ instead of $\mbpole$, we find
to five loop order,\cite{BBZ}
\bea
  \Gamma&=&{G_F^2|V_{ub}|^2\over192\pi^3}\,m_B^5
  \Bigg[1-2.41{\alpha_s\over\pi}
  -2.98\left({\alpha_s\over\pi}\right)^2\beta_0
  -4.43\left({\alpha_s\over\pi}\right)^3\beta_0^2\nonumber\\
  &&\qquad\qquad\qquad\mbox{}
  -7.67\left({\alpha_s\over\pi}\right)^4\beta_0^3
  -15.7\left({\alpha_s\over\pi}\right)^5\beta_0^4+\dots
  -5{\bar\Lambda\over m_B}+\dots\Bigg]\nonumber\\
  &=&{G_F^2|V_{ub}|^2\over192\pi^3}\,m_B^5\,\Big[
  1-0.061-0.120-0.107-0.111-0.136+\dots\nonumber\\
  &&\qquad\qquad\qquad\mbox{}-5\bar\Lambda/m_B+\dots\Big]\,,
\eea
for $\alpha_s(m_b)=0.21$ and $\beta_0=9$.  As we see, not
only does the perturbation series fail to converge, it does
not even have an apparent smallest term, where one should
truncate to minimize the error of the asymptotic series.  The
series for $\langle s_H\rangle$ exhibits a similar
behavior,\cite{FLS96}
\bea
  \langle s_H\rangle&=&m_B^2\Bigg[0.20{\alpha_s\over\pi}
  +0.35\left({\alpha_s\over\pi}\right)^2\beta_0
  +0.64\left({\alpha_s\over\pi}\right)^3\beta_0^2
  +1.29\left({\alpha_s\over\pi}\right)^4\beta_0^3\nonumber\\
  &&\qquad\qquad\qquad\mbox{}
  +2.95\left({\alpha_s\over\pi}\right)^5\beta_0^4+\dots
  +{7\over10}{\bar\Lambda\over m_B}+\dots\Bigg]\nonumber\\
  &=&m_B^2\,\big[
  0.0135+0.0141+0.0156+0.0189+0.0261+\dots\nonumber\\
  &&\qquad\qquad\qquad\mbox{}-7\bar\Lambda/10m_B
  +\dots\big]\,.
\eea
However, the situation improves dramatically if we eliminate
$\bar\Lambda$ and write $\Gamma$ directly in terms of
$\langle s_H\rangle$,
\bea
  \Gamma&=&{G_F^2|V_{ub}|^2\over192\pi^3}\,m_B^5\bigg[
  1-7.14{\langle s_H\rangle\over m_B^2}\nonumber\\
  &&\qquad\qquad\mbox{}-0.064
  -0.020-0.0002-0.022-0.047+\dots
  \bigg].
\eea
By truncating this series at its smallest term, 0.0002, we
obtain a new expression in which the theoretical errors are
under control.  The price is that we must now measure a
second quantity, $\langle s_H\rangle$, in the same decay.

\begin{figure}
\epsfysize10cm
\hfil\epsfbox{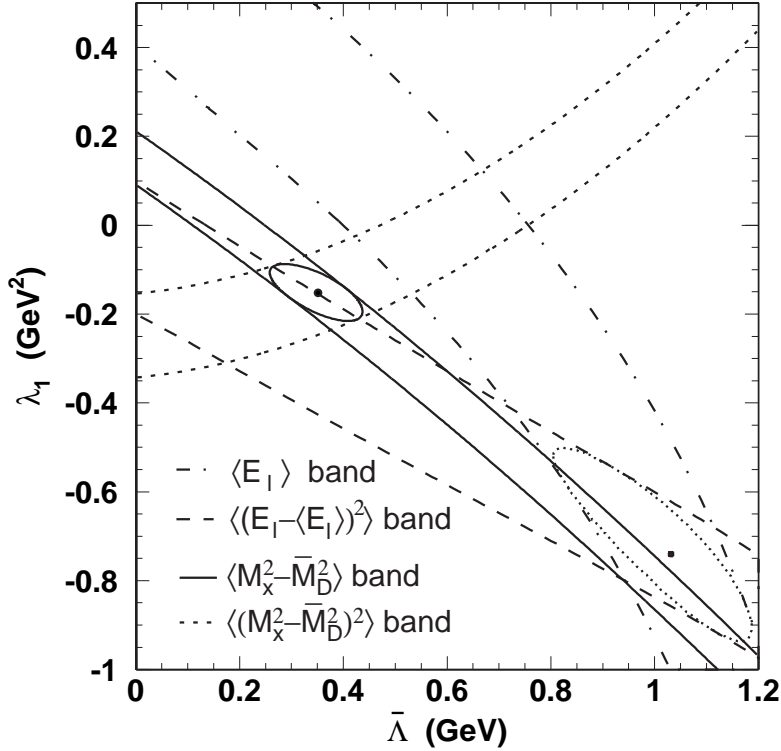}\hfill
\caption{The CLEO determination of $\bar\Lambda$ and
$\lambda_1$ from $B\to X_c\ell\bar\nu$.}
\label{fig:cleo}
\end{figure}

In principle, the same procedure works for decays to
charm.\cite{FLS96}  In practice, it is best to combine a
number of determinations of $\bar\Lambda$.  This has been
done by CLEO,\cite{CLEO2} which has performed a comparison
of measurements of the moments $\langle E_\ell\rangle$,
$\langle E_\ell^2\rangle$, $\langle
s_H-\overline m_D^2\rangle$ and $\langle
(s_H-\overline m_D^2)^2\rangle$.  The results, reproduced in
Fig.~\ref{fig:cleo}, are somewhat disappointing, in that one
does not obtain a very consistent determination of
$\bar\Lambda$ and $\lambda_1$.  Perhaps this is just a
fluctuation, or perhaps this is a sign that parton-hadron
duality is failing in these days.

An alternative approach is to express the width $\Gamma$ in
terms of the running mass $\overline m_b(m_b)$ instead of
another inclusive observable.\cite{Bigi,BBB}  Since the
$\overline{\rm MS}$ mass is a short distance quantity, this
also eliminates the infrared renormalon, which is associated
with long distance physics.  However, from a
phenomenological point of view, it raises the question of how
the running mass is to be determined from experiment. 
Possibilities include quarkonium spectroscopy, QCD sum rules,
and lattice calculations, but in all of these cases it is
important to determine reliably the accuracy of the
method, and how to deal with renormalon ambiguities in a
manner that is {\it consistent\/} with their treatment in the
calculation of $\Gamma$.  Nevertheless, such an approach,
particularly one based on analyzing quarkonium and
production near threshold, should eventually prove
fruitful.\cite{Beneke}

\subsection{Phenomenology of $V_{cb}$ and $V_{ub}$}

Despite this ambiguous situation, groups have presented
extractions of $V_{cb}$ based on inclusive semileptonic $b$
decays by simply inserting ``reasonable'' values for
$\bar\Lambda$ and $\lambda_1$.  Such an approach clearly
has its dangers!  At any rate, the quoted result
is~\cite{Poling}
\be
  |V_{cb}|=(40.0\pm0.4\pm2.4)\times10^{-3}\,.
\ee
The lack of controversy about this procedure is no doubt
due in part to the fact that this number is quite consistent
with that determined from the analysis of exclusive decays.

There are additional, much more interesting, problems with
extracting $|V_{ub}|$ from the inclusive decay $B\to
X_u\ell\bar\nu$.  They arise from the problem that the
process $B\to X_c\ell\bar\nu$ presents an overwhelming
background to $B\to X_u\ell\bar\nu$.  The only way to avoid
this background is to restrict oneself to a corner of phase
space in which charmed final states are kinematically
inaccessible.  Existing experimental analyses isolate the
charmless decays by imposing the requirement
$E_\ell>(m_B^2-m_D^2)/2m_B$ or $s_H<m_D^2$.  Unfortunately,
the OPE can be shown to break down in these restricted
corners of the phase space.\cite{shape,hadmass,endpoint}

I do not have space here to explore this issue in much
detail, but it is easy to appreciate the essence of the
problem.  For massless final states, the OPE is an expansion
in powers of the light quark propagator,
$1/m_b(1-v\cdot\hat q+\hat q^2)$.  Over most of the final
state phase space, the denominator is of order
$m_b$ and the OPE is well behaved.  But there exist
configurations for which both the denominator vanishes and the
operator matrix elements which appear in the OPE are
nonzero.  It turns out that the dangerous region is when
$v\cdot\hat q\to{1\over2}$ and $\hat q^2\to0$.  The precise
form of the divergence depends on the kinematic distribution
being studied.  The general form, however, is universal.  In
this ``endpoint region'', let $y$ be a scaled variable such that
$y\to1$ at the kinematic endpoint.  For example, for
${\rm d}\Gamma/{\rm d}E_\ell$, we take $y=2E_\ell/m_b$, and
for ${\rm d}\Gamma/{\rm d}s_H$ we take $y=s_H/\bar\Lambda
m_b$.  Then near $y=1$, the OPE takes the general form
\be\label{badope}
  {{\rm d}\Gamma\over{\rm d}y}\propto\sum_{n=0}^\infty
  c_n{A_n\over m_b^n(1-y)^n}\,,
\ee
where $c_n$ are coefficients of order one, and the $A_n$ are
moments defined by 
\be
  \langle B(v)|\, \bar h_v\, iD_{\mu_1} \ldots
  iD_{\mu_n}\, h_v\,
 |B(v)\rangle/2m_B = A_n\, v_{\mu_1} \ldots v_{\mu_n}
  +\dots\,.
\ee
The ellipses represent terms involving factors of the
metric tensor $g_{\mu_i\mu_j}$, which are subleading.  Since
the $A_n$ are associated with totally symmetric combinations
of the covariant derivatives, they may be interpreted
roughly as the moments of the heavy quark momentum.  Note
that as defined, $A_0=1$, $A_1=0$ and $A_2=\lambda_1$.

The divergences in Eq.~(\ref{badope}) can be controlled only
if one integrates over a large enough region near the
endpoint, $1-\delta\le y\le1$.  For
$\delta\sim\bar\Lambda/m_b$, one finds a series for $\Gamma$
which does not converge, since the individual terms are of
order $A_n/\bar\Lambda^n\sim1$.  This situation reflects a
dependence of the shape of ${{\rm d}\Gamma/{\rm d}y}$ on the
{\it entire\/} $b$ quark momentum distribution in the $B$
meson.  Since, for example, the window $2.3\,{\rm
GeV}<E_\ell<2.6\,{\rm GeV}$ corresponds to
$\delta\simeq0.1$, this problem pollutes the extraction of
$|V_{ub}|$ from ${{\rm d}\Gamma/{\rm d}E_\ell}$.  One is
forced to introduce a model for the $b$ wavefunction, with
the attendant uncontrolled theoretical uncertainties.  The
same is true, it turns out, for ${\rm d}\Gamma/{\rm d}s_H$. 
The current ``best'' measurement of $V_{ub}$ from LEP, based
on an inclusive analysis, is~\cite{LEPVub99}
\be
  |V_{ub}|=\left[4.05{}^{+0.39}_{-0.46}({\rm stat.})
  {}^{+0.43}_{-0.51}{}^{+0.23}_{-0.27}({\rm sys.})
  \pm0.02(\tau_b)\pm0.16({\rm HQE})\right]\times10^{-3}\,,
\ee
or approximately
$|V_{ub}/V_{cb}|=0.104^{+0.015}_{-0.018}$. 
While these analyses are experimentally very sophisticated,
they rely intensively on a two-parameter model of the $b$
quark wavefunction.  Essentially, in such a parameterization
all moments of the $b$ momentum distribution are correlated
with the first two nonzero ones, a constraint which is
unphysical.  Even if the two parameters are varied within
``reasonable'' ranges, it is doubtful that such a restrictive
choice of model captures reliably the true uncertainty in
$|V_{ub}|$ from our ignorance of the structure of the $B$
meson.  While the central value which is obtained in these
analyses is reasonable, the realistic theoretical error
which should be assigned is not yet well understood.

A recent analysis by CLEO of the exclusive decay
$\bar B\to\rho\ell\,\bar\nu$ yields~\cite{Behrens:1999vv}
\be
  |V_{ub}|=\left[3.25\pm0.14({\rm stat.})
  {}^{+0.21}_{-0.29}({\rm syst.})\pm0.55({\rm
  {\rm theory}})\right]\times10^{-3}\,,
\ee
or approximately $|V_{ub}/V_{cb}|=0.083^{+0.015}_{-0.016}$,
essentially consistent with the LEP
result.  In this case the reliance on
models is quite explicit, since one needs the hadronic form
factor $\langle\rho|\,\bar
u\gamma^\mu(1-\gamma^5)b\,|\bar B\rangle$ over the range of
momentum transfer to the leptons.  The CLEO
measurement relies on models based on QCD sum rules, which have
uncertainties which are hard to quantify.  Hence, just as in
the case of the LEP measurement, the quoted errors should not
be taken terribly seriously.  All of the current constraints
are consistent with
$|V_{ub}/V_{cb}|=0.090\pm0.025$, where I strongly prefer this
more conservative estimate of the theoretical errors. The
problem lies not in the experimental analyses, but in our
insufficient understanding of hadron~dynamics.

Recently it has been pointed out~\cite{BLL} that the one may
alternatively reject the charm background by studying the
distribution ${\rm d}\Gamma/{\rm d}q^2$ and restricting
oneself to  $q^2>(m_B-m_D)^2$.  Not only does this cut
eliminate charmed final states, but it also avoids the
troublesome region near $q^2=0$.  Hence such a determination
would not be polluted by the divergence described above. 
Whether the neutrino reconstruction algorithms of the $B$
Factories will be up to the task of this measurement yet
remains to be seen.

\section{Concluding Remarks}

Unfortunately, we have had time in these lectures only to
introduce a very few of the many applications of heavy quark
symmetry and the HQET to the physics of heavy hadrons.  Since
its development less than ten years ago, it has become one of
the basic tools of QCD phenomenology.  Much of the popularity
and utility of the HQET certainly come from its essential
simplicity.  The elementary observation that the physics of
heavy hadrons can be divided into interactions characterized
by short and long distances gives us immediately a clear and
compelling intuition for the properties of heavy-light
systems.  The straightforward manipulations which lead to the
HQET then allow this intuition to form the basis for a new
systematic expansion of QCD.  The deeper understanding of
heavy hadrons which we thereby obtain is
increasingly important as the $B$ Factory Era begins.

\section*{Acknowledgements}
It is a pleasure to thank the organizers of TASI-2000
for the opportunity to present these lectures, and for
arranging a most interesting and pleasant summer school. 
This work was supported by the National Science Foundation
under Grant No.~PHY-9404057, by the Department of Energy
under Outstanding Junior Investigator Award
No.~DE-FG02-94ER40869, and by the Research Corporation
under the Cottrell Scholarship program.

\end{document}